\documentclass[12pt]{article}

\usepackage{a4,amsmath,amssymb,amsthm,amscd,cite,graphicx}
\usepackage{hyperref} \hypersetup{colorlinks, linkcolor=red }
\usepackage{tikz}
\usetikzlibrary{intersections,calc,matrix,arrows, decorations.markings,shapes}
\usepackage{multirow}
\newtheorem{theorem}{Theorem}[section]
\newtheorem{mydef}[theorem]{Definition}

\def\Bbb{\mathbb} \def\BZ{\Bbb Z} \def\BR{\Bbb R} 
  \def\BH{\mathbb{H}}

\newcommand{\slhat}{{\widehat{sl(2)}}}

\catcode`@=11 \@addtoreset{equation}{section} \catcode`@=12

\begin{document}

\begin{titlepage}
\begin{flushright}
 July 2022\\
Version 1.2
\texttt{arXiv:2207.10502 [hep-th]}
\end{flushright}
\begin{center}
\textsf{\large $\slhat$ decomposition of denominator formulae of some BKM Lie superalgebras - II }\\[12pt]
Suresh Govindarajan$^{a}$ and Mohammad Shabbir$^{b,\,c}$ \\[4pt]
$^a$Department of Physics,
Indian Institute of Technology Madras,
Chennai 600036 India\\
 $^b$ The Institute of Mathematical Sciences, 
CIT Campus, Taramani
Chennai 600113, India \\
$^c$
Homi Bhabha National Institute
Training School Complex, Anushakti Nagar
Mumbai 400085, India
\\[4pt]
Email: suresh@physics.iitm.ac.in, mshabbir@imsc.res.in
\end{center}
\begin{abstract}
The square-root of Siegel modular forms of CHL $\BZ_N$ orbifolds of type II compactifications are denominator formulae for some Borcherds-Kac-Moody Lie superalgebras for $N=1,2,3,4$. We study the decomposition  of these Siegel modular forms in terms of characters of two sub-algebras: one is a $\slhat$ and the second is a Borcherds extension of the $\slhat$.  This is a continuation of our previous work where we studied the case of Siegel modular forms appearing in the context of Umbral moonshine. This situation is more intricate and provides us with a new example (for $N=5$) that did not appear in that case. We restrict our analysis to the first $N$ terms in the expansion as a first attempt at deconstructing the Siegel modular forms and unravelling the structure of potentially new Lie algebras that occur for $N=5,6$.
\end{abstract}
\end{titlepage}

\section{Introduction}

In this work, we continue the study of Siegel modular forms that are, in some cases, the denominator formulae for some Borcherds-Kac-Moody (BKM) Lie superalgebras. These Siegel modular forms include examples for which the Lie algebra connection is not yet known. For such examples, the eventual goal is to prove (or disprove) the existence of Lie algebras whose denominator formulae are given by these Siegel modular forms. 

  In our previous work\cite{Govindarajan:2021pkk}, we studied a family of Siegel modular forms that are associated with Umbral moonshine\cite{Cheng:2012tq}. Here we consider Siegel modular forms that are associated with $L_2(11)$-moonshine\cite{Govindarajan:2018ted,Govindarajan:2019ezd}. The squares of these Siegel modular forms are the generating function of quarter BPS states in CHL $\BZ_N$ orbifolds (for $N=1,2,\ldots,6$)\cite{Dijkgraaf:1996it,Cheng:2008kt,Govindarajan:2009qt,Govindarajan:2019ezd}. The main tool to probe the structure of the Lie algebras are  two subalgebras: one is a $\slhat$ subalgebra and the other is a Borcherds extension of the $\slhat$ subalgebra. We rewrite the Siegel modular forms in terms of characters of the sub-algebras -- it enables us to cleanly track simple roots that appear in the denominator formulae.

 For simplicity, we focus on the situations when $N$ is prime, i.e., $N=2,3,5$.  These are modular forms of weight $k(N)+1=12/(N+1)$ of a level $N$ subgroup of $Sp(4,\BZ)$. The connection with Mathieu and $L_{2}(11)$ moonshine leads to a product formula given in Eq. \eqref{Productside}, for the Siegel modular forms\cite{Govindarajan:2011em,Persson:2013xpa,Govindarajan:2018ted}. For the prime cases, it is consistent with the product formulae given by David et al.\cite{David:2006ji} in the context of dyon counting. We rewrite the Siegel modular form  as follows:
\begin{equation}
\Delta^{(N)}_{k(N)}(\mathbf{Z}) = s^{1/2}\,\phi^{(N)}_{k(N),1/2}(\tau,z) \times \Big[ 1 + \sum_{m=1}^\infty s^m\, \Psi^{(N)}_{0,m}(\tau,z).)\Big]\ ,
\end{equation}
The Jacobi forms $\Psi^{(N)}_{0,m}(\tau,z)$ will be the main object of our study. They are Jacobi forms of the congruence subgroup $\Gamma^0(N)$ with weight zero and index $m$. We obtain explicit formulae for these Jacobi forms in terms of standard modular forms for $m\leq N$. The analogous expansion in our previous work\cite{Govindarajan:2021pkk} had non-vanishing terms only for indices that were multiples of $N$. 

We wish to show that the Siegel modular forms $\Delta^{(N)}_{k(N)}(\mathbf{Z})$ are extensions of the Kac-Moody Lie algebra $\mathfrak{g}(A^{(N)})$ obtained from the Cartan matrix, $A^{(N)}$ , defined in Eq. \eqref{CartanMatrix}. We call the extension $\mathcal{B}_N^{CHL}(A^{(N)})$ -- the $CHL$ refers to the fact that the square of the modular forms are the generating functions of quarter BPS states in $CHL$ $\BZ_N$ orbifolds\cite{Jatkar:2005bh,Govindarajan:2009qt,Cheng:2008kt}. The Cartan matrices $A^{(N)}$ are obtained from the walls of marginal stability in these models\cite{Sen:2007vb}. These have nice behaviour only for $N=1,2,\ldots,6$. The expectation is that for $N\leq 4$, the extension $\mathcal{B}_N^{CHL}(A^{(N)})$ is the usual Borcherds extension of $\mathfrak{g}(A^{(N)})$ which leads to the sum side of the denominator formula given in Eq. \eqref{Bextension}. The Borcherds correction term is shown symbolically as $T$ in this formula -- it is the contribution that one obtains by adding imaginary simple roots i.e., roots with negative or zero norm.

A Cartan matrix can also be obtained as the matrix of inner products of simple root vectors which generate a root lattice. In all the six examples, the Cartan matrix has rank three and the root lattice is in Lorentzian space $\mathbb{R}^{2,1}$. A special feature of these lattices is that they admit a lattice Weyl vector $\varrho^{(N)}$ with inner product $\langle \varrho^{(N)}, \alpha\rangle =-1$ where $\alpha$ is a simple root. Such lattices have been studied by Nikulin and the corresponding Lie algebra connection by Gritsenko and Nikulin\cite{Gritsenko:2002}. An important result from Gritsenko and Nikulin is that the cases of $N\leq 4$ in our examples can admit Borcherds extensions. This is why we expect that  $\mathcal{B}_N^{CHL}(A^{(N)})$ are Borcherds extensions. Unlike the examples considered in our previous work\cite{Govindarajan:2021pkk}, we are unaware of a proof that this is indeed the case for $N\leq 4$.

The reason one hopes that there might be a Lie algebra for $N=5,6$ is a physical one. The dyon counting generating function provides us with Siegel modular forms that transform covariantly under the Weyl group of $\mathfrak{g}(A^{(N)})$. We have three examples of this variety, one of which was considered in \cite{Govindarajan:2021pkk}. We restrict to the case of $N=5$ for simplicity in this work commenting on some aspects of the $N=6$ example. Our goal in this work is a modest one. We study two sub-algebras of $\mathcal{B}_N^{CHL}(A^{(N)})$, one is an $\slhat\in\mathfrak{g}(A^{(N)} $ and another is a Borcherds extension of the $\slhat$ that we call $\mathcal{B}_N^{CHL}(\slhat)$. Interestingly, these subalgebras are the best examples to understand the idea behind the Borcherds extension. The positive roots of the Lie algebra $\mathcal{B}_N^{CHL}(\slhat)$ that are not in the sub-algebra will organise into a representation of the sub-algebra. 
This is the motivation for us to look into character decompositions of the 
$\Psi_{0,m}^{(N)}(\tau,z)$  in terms of $\slhat$ and   $\mathcal{B}_N^{CHL}(\slhat)$. 

The goal of the present paper is a modest one. We would like to understand the structure of the irreducible roots that appear in the first $N$ terms. The main result of this paper is that we are able to characterize all the roots that appear to this order and they are consistent with our expectations. There are some surprises. For instance, $\Psi_{0,2}^{(3)}(\tau,z)$ vanishes. This is due to perfect cancellations between two different terms. We see the appearance of a real simple fermionic root in the $N=5$ example which has some peculiar properties. This is the first term that does not appear as a Borcherds extension. This is consistent with a  no-go theorem of Gritsenko and Nikulin that suggests that modifications be needed for the cases of $A^{(5)}$ and $A^{(6)}$\cite{Gritsenko:2002}. 

The organization of the paper is as follows. The introductory section is followed by section 2 where we provide the Lie algebra background as well as develop the notation used in the rest of the paper. Section 3 is where we obtain vector-valued modular forms(vvmf) of $\Gamma^0(N)$ by expanding in terms of $\slhat$ and $\mathcal{B}_N^{CHL}(\slhat)$. 
The Fourier coefficients of the vvmf can be identified with the multiplicities of roots that appear. We closely track all roots of non-negative norm. For $N=5$ this enables us to see the presence of a fermionic real simple root that does not fit a Borcherds extension.  In section 4, we convert the vvmfs of $\Gamma^0(N)$ into vvmfs of the full modular group. For one example alone, we are able to identify the vvmf to be a solution of a modular differential equation studied by Gannon\cite{Gannon:2013jua}. In all other situations, the rank of the vvmf is too large for us to numerically determine the modular differential equation. We conclude in section 5 with some remarks. An appendix is devoted to providing the background necessary for the computations that we have done in this paper.

\section{The Lie algebra background}

A  vector in $\BR^{2,1}$ can be represented by a real symmetric $2\times 2$ matrix\cite{Feingold1983,Cheng:2008fc}. 
\[
\begin{pmatrix}
x \\ y \\ t
\end{pmatrix}\longleftrightarrow v=\begin{pmatrix}
t + y & x \\ x & t-y
\end{pmatrix}
\]
with norm $\langle v,v \rangle=-2 \det(v)= 2(x^2 + y^2 -t^2)$. Consider the two vectors in given by
\begin{equation}
\alpha_1= \begin{pmatrix} 2 & 1 \\ 1 & 0 \end{pmatrix}\text{ and } \ 
\alpha_2=\begin{pmatrix} 0 & -1 \\ -1 & 0 \end{pmatrix} \ .
\end{equation}
Starting from these two \textit{root} vectors construct new root vectors as follows:
\begin{equation}
\alpha_{a+2m} = \left(\gamma^{(N)}\right)^m \cdot \alpha_a \cdot \left((\gamma^{(N)})^T\right)^m\text{ for } a=1,2,
\end{equation}
where $
\gamma^{(N)}=\left(\begin{matrix}
                1 & -1 \\  N & 1-N
               \end{matrix}  \right)
$. Note that $\gamma^{(N)}$ and $-\gamma^{(N)}$ have identical action on the $\alpha_i$. For $N\leq 3$, $\gamma^{(N)}$ has finite order and infinite order for $N>3$. 

Let $\mathbf{X}_N$ denote the ordered sequence of distinct root vectors $\alpha_i$ generated in this fashion. 
\begin{equation}
\mathbf{X}_N= (\alpha_i) \text{ for }i\in \mathcal{S}_N=\begin{cases}(1,2,3\text{ mod } 3)\ , & N=1 \\
(0,1,2,3\text{ mod } 4)\ , & N=2 \\
(0,1,2,3,4,5\text{ mod } 6)\ , & N=3 \\
\BZ \ , & N=4,5,6
\end{cases}\ .
\end{equation}
There is a Weyl vector $\varrho^{(N)}$
\begin{equation}
\varrho^{(N)}=\begin{pmatrix} 1/N & 1/2 \\ 1/2 & 1\end{pmatrix}\ ,
\end{equation}
with norm $\langle\varrho^{(N)},\varrho^{(N)}\rangle=(\frac12 -\frac2N)$ with $\langle\varrho^{(N)},\alpha\rangle= -1$ for all $\alpha\in \mathbf{X}_N$.
 
Let $A^{(N)}$ for $N=1,2,\ldots,6$ denote  matrices given by the Gram matrix of the root vectors $\mathbf{X}_N$
\begin{equation}\label{CartanMatrix}
A^{(N)}= (a_{nm}):= \langle \alpha_m,\alpha_n\rangle\ . 
\end{equation}
One has $a_{nm}= 2 - \frac{4}{N-4}(\lambda_N^{n-m} + \lambda_N^{m-n}-2)$,
where $\lambda_N$ is any solution of the quadratic equation 
\begin{equation*}
\lambda^2 -(N-2)\lambda + 1 =0\ .
\end{equation*}

Let $\mathfrak{g}(A^{(N)})$ denote the Kac-Moody algebra associated with the Cartan matrix $A^{(N)}$\cite{Kac1990}. 
Recall that the Kac-Moody algebra, $\mathfrak{g}(A)$, associated with a Cartan matrix $A=(a_{mn})$ (with $m,n\in I$) is given by the generators $(e_m,h_m, f_m)$ with Lie brackets
\[
[e_m,f_n]=\delta_{mn}\ h_m\ , \ [h_m,e_n]= a_{mn}\ e_n\ , \ [h_m,f_n]=-a_{mn}\ f_n\ , \ [h_m,h_n]=0\ ,
\]
subject to the Serre relations
\[
(\text{ad}\ e_m)^{-a_{mn}+1} e_n=0\ , \ (\text{ad}\ f_m)^{-a_{mn}+1} f_n=0\quad m\neq n\ ,
\]
where $(\text{ad} x) y =[x,y]$.

The Borcherds extension of a Kac-Moody algebra, a BKM Lie algebra,  is obtained by adding imaginary simple roots to $\mathfrak{g}(A^{(N)})$.  A simple description is given by considering the Weyl denominator formula which takes the form:
 \begin{equation}\label{Bextension}
 \Delta=\sum_{w\in W}\text{det}(w) w \Big[T\ e^{-\varrho}\Big]  = e^{-\varrho}\ \prod_{\alpha\in L_+} (1-e^{-\alpha})^{\text{mult}(\alpha)} \ .
\end{equation}
In the above formula, $W$ is the Weyl group generated by elementary reflections due to simple roots, $\varrho$ is the Weyl vector, $L_+$ is the set of positive roots and $\text{mult}(\alpha)$ is the multiplicity of the root $\alpha$. The case when $T=1$ is for the case of Kac-Moody algebras. $T$ is the Borcherds correction term that takes into account the presence imaginary simple roots. (See appendix B of \cite{Govindarajan:2021pkk} and references therein for a detailed description.) A key aspect of the Borcherds extension is that $\Delta$ is a suitable automorphic form that admits a product formula.

\noindent \textbf{An example:} Let $A=\begin{pmatrix}
2 & -2 \\ -2 & 2
\end{pmatrix}$. Then, $\mathfrak{g}(A)$ is the $\slhat$ Kac-Moody Lie algebra with simple roots $(\alpha_1,\alpha_2)$ and $\delta=\alpha_1+\alpha_2$ is an imaginary root with zero root. We will consider a family of Borcherds corrections that appear in this work.
 For $N=1,2,3,5$, consider a situation where has $12/(N+1)$ distinct imaginary simple roots of weight $\tfrac1N(\delta,2\delta,3\delta,\ldots)$ and $(12/(N+1))-3$ imaginary simple roots of weight $(\delta,2\delta,3\delta,\ldots)$. The Borcherds correction factor due to these imaginary simple roots takes the form
\[
T_N(\delta) = \prod_{j=1}^{\infty} \left(1-e^{-\tfrac{j\delta}N}\right)^{\tfrac{12}{N+1}}\left(1-e^{-j\delta}\right)^{-3+\tfrac{12}{N+1}}\ .
\]
For $N=5$ a negative power appears in the second term in the infinite product. The imaginary simple roots in this case correspond to isotropic fermionic simple roots and we consider a superdenominator formula to account for this.
Identifying $e^{-\delta}\sim q=\exp(2\pi i\tau)$, we obtain a function of $\tau$. Let
\begin{equation}
T_N(\tau) = \prod_{j=1}^{\infty} \left(1-q^{j/N}\right)^{\tfrac{12}{N+1}}\left(1-q^{j}\right)^{-3+\tfrac{12}{N+1}}\ .
\end{equation}
Up to an overall power of $q$, $T_N(\tau)$ can be expressed in terms of products of the Dedekind eta function.
 The automorphic form, that is denoted by $\Delta$ in Eq. \eqref{Bextension}, for these examples is given by the Jacobi form $\phi_{k(N),1/2}(\tau,z)$ defined in Eq. \eqref{addseed}. We will refer to these Borcherds-Kac-Moody Lie algebras by $\mathcal{B}_N^{CHL}(\slhat)$. As can be seen, there can be several inequivalent Borcherds extensions of a Kac-Moody Lie algebra. 

\subsection{Embedding $\slhat$ in $\mathfrak{g}(A^{(N)})$}

The Cartan matrices. $A^{(N)}$ considered in paper I {\cite{Govindarajan:2021pkk} are identical to the ones that appear here as well. Thus, the embedding of $\slhat$ into $\mathfrak{g}(A^{(N)})$ works here as well. Let $(e,h,f)$ be the generators of $sl(2)$.
 The affine Lie algebra $\widehat{sl(2)}$ is defined by
 \[
 \widehat{sl(2)} = sl(2)\otimes\mathbb{C}[t,t^{-1}]\oplus \mathbb{C}\,\hat{k} \oplus \mathbb{C}\, d\ ,
 \]
 where $\hat{k}$ is the central extension and $d=-t d/dt$ is the derivation.

 We identify the Lie subalgebra of $\mathfrak{g}(A^{(N)})$ generated by $e_1, f_1, e_2, f_2, h_1, h_2$ and $h_3$ with  $\widehat{sl(2)}$ Lie algebra. We choose the identification similar to the one considered by  Feingold and Frenkel\cite{Feingold1983}.
 \[
 e\otimes 1 =e_2\ ,\  f\otimes 1=f_2\ ,\  f \otimes t =e_1\ , \  e\otimes t^{-1} =f_1 \ ,
 \]
 For the Cartan subalgebra of $\widehat{sl(2)}$, using the above identification, we obtain
 \[
 h_1 =-h\otimes 1 +\hat{k}\ , \ h_2 = h \otimes 1 \ ,\ h_3= - h\otimes 1  +4N\, d\ .
 \]
The inverse is
\[
h\otimes 1 = h_2\ , \  \hat{k} = h_1 + h_2 \ , \ d =  \frac{1}{4N} (h_2 + h_3) \ .
\]

\subsection{The $\mathcal{B}^{CHL}(A^{(N)})$ Lie algebras}

Let $\mathcal{B}^{CHL}(A^{(N)})$ denote an extension of the $\mathfrak{g}(A^{(N)})$ whose denominator formula is given by the Siegel modular forms, $\Delta^{(N)}_{k(N)}(\textbf{Z})$  which we define next\footnote{Here $\mathbf{Z}=\begin{pmatrix}
\tau & z \\ z & \tau'
\end{pmatrix}$ is a point in the Siegel upper half space,  $\mathbb{H}_2$. See appendix A.3.}.  Then the BKM Lie algebras $\mathcal{B}_N^{CHL}(\slhat)$ are naturally sub-algebras of $\mathcal{B}^{CHL}(A^{(N)})$. 

A connection with Mathieu and $L_{2}(11)$ moonshine leads to the following formula for a Siegel modular form\cite{Govindarajan:2011em,Persson:2013xpa,Govindarajan:2018ted}. Let $g\in L_2(11)_B$  be an element of order $N\leq6$. A second-quantized version of moonshine gives the following formula for 
$\Delta^{(N)}_{k(N)}(\textbf{Z})$.
\begin{equation}\label{smfdef}
\Delta^{(N)}_{k(N)}(\textbf{Z}) =s^{1/2} \phi_{k(N),1/2}(\tau,z)\exp\left[- \frac{1}{m}\sum\limits_{m=1}^{\infty}s^m \psi_{0,1}^{[1,g]}(\tau,z)\Big|T(m)\right]
\end{equation}
where the Hecke-like operator $T(m)$ is defined as follows\footnote{Here $ \psi _{0,1}^{(N)[g^s,g^r]}$ is half the $g^r$-twisted elliptic genus of $K3$ twined by the element $g^s$. In other words, the trace is over the Hilbert space twisted by $g^r$ with insertion of $g^s$ (`twined') in the trace.}
\[
\psi_{0,1}^{(N)[1,g]}(\tau,z)\Big|T(m) := \frac1m\sum\limits_{ad=m} \sum\limits_{b=0}^{d-1} \psi _{0,1}^{(N)[g^{-b},g]}\left(\tfrac{a \tau+b}{d} ,az \right)
\]
and 
\begin{align}\label{addseed}
\phi_{k(N),1/2}(\tau,z)=\frac{\theta_1(\tau,z)}{\eta(\tau)^3}\
\eta^{[1,g]}(\tau)
\end{align}
are index half Jacobi forms with the eta products $\eta^{[1,g]}(\tau)$ defined in Table \ref{etaproducts}. It has been shown in ref. \cite{Govindarajan:2019ezd} that this leads to a Borcherds-type product formula for $\Delta^{(N)}_{k(N)}(\textbf{Z})$.
Consider the Fourier expansion
\begin{equation}
\psi^{[g^{b},g^d]}_{0,1}(\tau,z)=\sum_{n\in\mathbb{Z}, n\ge0}\sum_{\ell \in \mathbb{Z}} c^{[b,d]}(n,\ell)\ q^{\frac{n}{N}} r^\ell\ ,
\end{equation}
where  $g$ is of order $N$, $q=e^{2\pi i\tau}$ and $r=e^{2\pi i z}$.   Define $\tilde{c}^{[\alpha,d]}(n,\ell)$ as follows
(with $\omega_N=\exp(2\pi i/N)$)
\begin{align}
\tilde{c}^{[\alpha,d]}(n,\ell)=\frac1N\sum_{b =0}^{N-1} \ (\omega_N)^{-\alpha b} \ c^{[b,d]}(n,\ell)\ .
\end{align}
Then one has the product formula that is provides the product side of the denominator formula that defines $\mathcal{B}^{CHL}(A^{(N)})$.
\begin{align}\label{Productside}
\Delta_{k(N)}^{(N)}(\mathbf{Z})=q^{1/2N} r^{1/2} s^{1/2} \times\prod_{m=0}^\infty\prod_{\alpha=0}^{N-1}\prod_{\substack{n\in \mathbb{Z}+\frac{\alpha}{N}\\ n\ge0}}\prod_{\ell\in \mathbb{Z}}(1-q^{n} r^
\ell s^m)^{ {\tilde{c}}^{[\alpha,m]}(nmN,\ell)}\ ,
\end{align}
with $s=e^{2\pi i\tau'}$.
 The modularity of the above formula is not manifest. However, it follows from a result in ref. \cite{Govindarajan:2020owu} that it is a Siegel modular form of a level $N$ subgroup of $Sp(4,\BZ)$.
 
The sum side of the Weyl denominator formula is usually obtained from an additive lift. There is a construction of Cl\'ery and Gritsenko that leads to closely related Siegel modular form (at level $N$) starting from a index half Jacobi form\cite{Clery2008}. It has been shown in \cite{Govindarajan:2018ted} that the expansion of this Siegel modular form about another cusp (given by the S-transform) matches with the product formula given in Eq. \eqref{Productside} to fairly high order. Combined with modularity, it is enough to prove that the two formulae are equivalent. It is not a clean formula in the sense that a closed formula was not given but the transformation rules for the Hecke operator were worked out on a case-by-case basis.

\begin{table}\label{etaproducts}
\centering
\begin{tabular}[hbt]{c|c|c|c|c}
 $N$ & 1 & 2& 3 & 5 \\
 \hline
Cycle shape  & $1^{12}$  & $1^4 2^4$ & $1^3 3^3$ & $1^2 5^2$ \phantom{\Big|} \\
\hline
$k(N)$ & 5 & 3 & 2 & 1 \\ \hline
 $\eta^{[1,g]}(\tau)$ & $\eta(\tau)^{12}$  & $\eta(\tau)^{4}\eta(\tau/2)^{4}$ & $\eta(\tau)^{3}\eta(\tau/3)^{3}$ & $\eta(\tau)^{2}\eta(\tau/5)^{2}$\phantom{\Big|} 
\end{tabular}.
\caption{Eta products}
\end{table}

\subsection{Covariance under the extended Weyl group}

The extended Weyl group of the root system $\mathbf{X}_N$ is generated by three types of generators\cite{Cheng:2008kt,Govindarajan:2009qt,Govindarajan:2019ezd}
\begin{enumerate}
\item The Weyl group $W$ of $\mathfrak{g}(A^{(N)})$ is generated by \textit{all} elementary Weyl reflections, $s_m$,  due to the simple roots $\alpha_m$ for all $m$ in $\mathcal{S}_N$, 
\item the generator $\gamma^{(N)}$, and
\item the generator $\widehat{\delta}=\begin{pmatrix}
-1 & 1 \\ 0 & 1
\end{pmatrix}$ which acts on roots via the action $\alpha \rightarrow 
\widehat{\delta}\cdot \alpha\cdot \widehat{\delta}^{\,T}$. It acts on the simple roots in $\mathbf{X}_N$ as an involution:
\[
\widehat{\delta}: \alpha_m \leftrightarrow \alpha_{3-m}\ .
\]
\end{enumerate}
The action of the generators of the extended Weyl group can be translated into an action on upper half space with coordinates $\mathbf{Z}$. With this in hand, one can show, using the modular properties of the Siegel modular forms, that
\[
\begin{split}
\Delta^{(N)}_{k(N)}(s_m\cdot \textbf{Z}) &= - \Delta^{(N)}_{k(N)}(\textbf{Z})\ , \\
\Delta^{(N)}_{k(N)}(\gamma^{(N)}\cdot \textbf{Z}) &= + \Delta^{(N)}_{k(N)}(\textbf{Z})
\ , \\
\Delta^{(N)}_{k(N)}(\widehat{\delta}\cdot \textbf{Z}) &= + \Delta^{(N)}_{k(N)}(\textbf{Z})\ .
\end{split}
\]
These properties show that the Siegel modular forms have the necessary covariance under the extended Weyl group.

\section{Deconstructing the Lie algebra}

The Siegel modular form defined in Eq. \eqref{smfdef} can be expanded as a power series in the variable $s$. The leading term in the expansion is $s^{1/2}\,\phi^{(N)}_{k(N),1/2}(\tau,z)$ which is the denominator formula for the sub-algebra $\mathcal{B}^{CHL}_N(\slhat)$.
\begin{equation}
\Delta^{(N)}_{k(N)}(\textbf{Z}) = s^{1/2}\,\phi^{(N)}_{k(N),1/2}(\tau,z) \Big[ 1 +\sum_{m=1}^\infty s^m\, \Psi^{(N)}_{0,m}(\tau,z)\Big]\ ,
\end{equation}
The above equations define the weight zero and index $m$ Jacobi forms $\Psi^{(N)}_{0,m}(\tau,z)$. Explicit formulae for the Jacobi forms can be obtained by expanding the exponential in Eq. \eqref{smfdef}. For instance, one obtains
\begin{align}
\Psi^{(N)}_{0,1}(\tau,z) &= -\psi_{0,1}^{(N)[1,g]}(\tau,z)\ , \\
\Psi^{(N)}_{0,2}(\tau,z) &= -\frac{1}{2}\left(\psi_{0,1}^{(N)[1,g]}(\tau,z)\Big|T(2)-(\psi_{0,1}^{(N)[1,g]}(\tau,z))^2\right) \ .
\end{align}
We will be studying  the first $N$ terms in the expansion. They can be rewritten in terms of standard modular forms thereby enabling us to have formulae that can be directly used. 
A weak Jacobi form of $\Gamma_0(N)$, $\xi_m$, of weight zero and index $m$ can be expanded as follows:
\[
\xi_m(\tau,z) = \sum_{j=0}^m  \alpha_j(\tau)\ A(\tau,z)^{m-j} B(\tau,z)^j\ ,
\]
where $\alpha_j(\tau)$ are weight $2j$ modular forms of $\Gamma_0(N)$ and $A(\tau,z)$, $B(\tau,z)$ are defined in Eq. \eqref{ABdef}. However the $ \Psi^{(N)}_{0,m}(\tau,z)$ are Jacobi forms of $\Gamma^0(N)$. Thus, we identify $\xi_m$ with their transform $ \Psi^{(N)}_{0,m}(\tau,z)|S$ as they are modular forms of $\Gamma_0(N)$. This method is useful as the generators of the ring of modular forms of $\Gamma_0(N)$ are well-known. We give the generators for the cases of interest in appendix \ref{ring}.

\subsection{Details of the examples}

We now present explicit formulae for the Jacobi forms $ \Psi^{(N)}_{0,m}(\tau,z)|S$ for $N=2,3,5$ and $m=1,\ldots,N$. 

\subsubsection{$N=2$}

The Weyl-Kac-Borcherds denominator formula is given by the weight three Siegel modular form of a level 2 subgroup of $Sp(4,\BZ)$.
\begin{equation}
\Delta^{(2)}_3(\mathbf{Z}) = s^{1/2}\,\phi^{(2)}_{3,1/2}(\tau,z) \Big[ 1 + s\, \Psi^{(2)}_{0,1}(\tau,z) + s^2\, \Psi^{(2)}_{0,2}(\tau,z) + O(s^3)\Big]\ ,
\end{equation}
where
\begin{align*}
\phi^{(2)}_{3,1/2}(\tau,z) &= \theta_1(\tau,z)\,\eta(\tau)^4 \eta(\tau/2)^4 \\
\Psi^{(2)}_{0,1}(\tau,z) & = \tfrac13 A(\tau,z) -\tfrac13 E_2^{(2)}(\tau/2)\, B(\tau,z)  \\
\Psi^{(2)}_{0,2}(\tau,z) &=-\tfrac1{72} A(\tau,z)^2 -\tfrac1{18} E_2^{(2)}(\tau/2) A(\tau,z) B(\tau,z) \\ 
&\quad + \left(\tfrac{29}{288} E_2^{(2)}(\tau/2)^2 -\tfrac1{32} E_4(\tau/2) \right) B(\tau,z)^2
\end{align*}
are Jacobi forms of $\Gamma^0(2)$. We expect to observe two real simple roots in 
$\Psi^{(2)}_{0,2}(\tau,z)$.

\subsubsection{$N=3$}
The Weyl-Kac-Borcherds denominator formula is given by the weight two Siegel modular form of a level 3 subgroup of $Sp(4,\BZ)$.
\begin{equation}
\Delta^{(3)}_2(\mathbf{Z}) = s^{1/2}\,\phi^{(3)}_{2,1/2}(\tau,z) \Big[ 1 + s\, \Psi^{(3)}_{0,1}(\tau,z) + s^2\, \Psi^{(3)}_{0,2}(\tau,z) + s^3\, \Psi^{(3)}_{0,3}(\tau,z) + O(s^4)\Big]\ ,
\end{equation}
where
\begin{align*}
\phi^{(3)}_{2,1/2}(\tau,z) &= \theta_1(\tau,z)\,\eta(\tau)^3 \eta(\tau/3)^3 \\
\Psi^{(2)}_{0,1}(\tau,z) & = \tfrac14 A(\tau,z) -\tfrac14 E_2^{(3)}(\tau/3)\, B(\tau,z)  \\
\Psi^{(3)}_{0,2}(\tau,z) &= 0\\
\Psi^{(3)}_{0,3}(\tau,z) &= \tfrac1{864} A(\tau,z)^3 -\tfrac1{96} E_2^{(3)}(\tau/3) A(\tau,z)^2 B(\tau,z) \\ 
& + \left(\tfrac{25}{1296} E_2^{(3)}(\tau/3)^2 -\tfrac{5}{2592} E_4(\tau/3) \right) A(\tau,z)B(\tau,z)^2  \\
&+ (-\tfrac{145}{11664}E_2^{(3)}(\tau/3)^3 +\tfrac{85}{23328}E_2^{(3)}(\tau/3)E_4(\tau/3) + \tfrac1{1458}E_6(\tau/3))B(\tau,z)^3
\end{align*}
are Jacobi forms of $\Gamma^0(3)$. It is interesting to observe that $\Psi^{(3)}_{0,2}(\tau,z) = 0$. This arises from a cancellation of multiple terms. The expectation is that there would have been no real simple roots and imaginary simple roots in this term. The vanishing says that there are no imaginary simple roots with negative norm. It could also be that there is a Bose-Fermi cancellation i.e., there are equal numbers of bosonic and fermionic roots. We expect to see two real simple roots in  $\Psi^{(3)}_{0,3}(\tau,z)$ which is non-vanishing.

\subsubsection{$N=5$}

The Weyl-Kac-Borcherds denominator formula is given by the weight one Siegel modular form of a level 5 subgroup of $Sp(4,\BZ)$.
\begin{equation}
\Delta^{(5)}_1(\mathbf{Z}) = s^{1/2}\,\phi^{(5)}_{1,1/2}(\tau,z) \Big[ 1 + \sum_{m=1}^5s^m\, \Psi^{(5)}_{0,m}(\tau,z)  + O(s^6)\Big]\ ,
\end{equation}
\begin{align*}
\phi^{(5)}_{1,1/2}(\tau,z) &= \theta_1(\tau,z)\,\eta(\tau/5)^2 \eta(\tau)^2 \\
\Psi^{(5)}_{0,1}(\tau,z) & = \tfrac15 A(\tau,z) -\tfrac15 E_2^{(5)}(\tau/5)\, B(\tau,z) 
\end{align*}
We have shortened $A(\tau,z), B(\tau,z)$ to $A,B$ to make equations more compact.
\begin{align*}
\Psi^{(5)}_{0,2}(\tau,z) &= -\tfrac1{144} A^2 -\tfrac1{72} E_2^{(5)}(\tau/5) A B \\ 
&\quad + \left(-\tfrac{53}{7200} E_2^{(5)}(\tau/5)^2 +\tfrac1{2400} E_4(\tau/5) -\tfrac{19}{200} \eta(\tau/5)^4\eta(\tau)^4 \right) B^2\\
\Psi^{(5)}_{0,3}(\tau,z) &= \tfrac{1}{864} A(\tau,z)^3 -\tfrac{1}{288} E_2^{(5)}(\tau/5) A^2 B \\ 
&\quad + \left(\tfrac{17}{4800} E_2^{(5)}(\tau/5)^2 -\tfrac{1}{14400} E_4(\tau/5)+\tfrac{19}{1200} \eta(\tau/5)^4\eta(\tau)^4 \right) A\,B^2\\
&+ \left(-\tfrac{53}{43200} E_2^{(5)}(\tau/5)^3 +\tfrac{1}{14400} E_2^{(5)}(\tau/5)E_4(\tau/5) -\tfrac{19}{1200} E_2^{(5)}(\tau/5) \eta(\tau/5)^4\eta(\tau)^4 \right) B^3\\
\Psi^{(5)}_{0,4}(\tau,z) 
&=  \tfrac{1}{20736} A^4  
-\tfrac{1}{5184} E_2^{(5)}(\tau/5) A^3\, B \\ 
&\quad + \left(\tfrac{17}{57600} E_2^{(5)}(\tau/5)^2 -\tfrac{1}{172800} E_4(\tau/5)+\tfrac{19}{14400} \eta(\tau/5)^4\eta(\tau)^4 \right) A^2\,B^2\\
&+ \left(-\tfrac{53}{259200} E_2^{(5)}(\tau/5)^3 +\tfrac{1}{86400} E_2^{(5)}(\tau/5)E_4(\tau/5) -\tfrac{19}{7200}  E_2^{(5)}(\tau/5) \eta(\tau/5)^4\eta(\tau)^4 \right) A\,B^3\\
&+ \left(\tfrac{2117}{25920000} E_2^{(5)}(\tau/5)^4 -\tfrac{1}{28800} E_2^{(5)}(\tau/5)^2E_4(\tau/5) +\tfrac{2641}{360000}  E_2^{(5)}(\tau/5)^2 \eta(\tau/5)^4\eta(\tau)^4 \right. \\
&\qquad\qquad \left. +\tfrac{11}{8640000} E_4(\tau/5)^2 + \tfrac{779}{60000}  \eta(\tau/5)^8\eta(\tau)^8\right) B^4
\end{align*}
are Jacobi forms of $\Gamma^0(5)$. We have not given an explicit formula for $\Psi^{(5)}_{0,5}(\tau,z)$ as the formula is big and unilluminating.

\subsection{Characters of $\slhat$ and $\mathcal{B}_N(\slhat)$ }

Consider the following roots
\begin{equation}
\alpha^{(N)}_0 =\begin{pmatrix}
2N-2 & 2N-1 \\ 2N-1 & 2N 
\end{pmatrix} \text{ and } \ 
\alpha^{(N)}_3= \begin{pmatrix} 0 & 1 \\ 1 & 2N \end{pmatrix} .
\end{equation}
We will track these real simple roots as well as the zero-norm imaginary simple roots 
\begin{equation}
\delta_N':=(\alpha^{(N)}_3+\alpha_2) \quad\text{and}\quad
 \delta_N'':=(\alpha^{(N)}_0+\alpha_2)\ .
\end{equation}
The subscript $N$ is to emphasise that they change with $N$ unlike the zero-norm imaginary simple root $\delta=(\alpha_1+\alpha_2)$.

The normalized $\widehat{sl(2)}$ character at level $k$, $\chi_{k,\ell}(\tau,z)$, is defined by
\begin{align}
\chi_{k,\ell}(\tau,z)&=\frac{\theta_{k+2,\ell+1}(\tau,z)-\theta_{k+2,-\ell-1}(\tau,z)}{\theta_{2,1}(\tau,z)-\theta_{2,-1}(\tau,z)} \text{ for } k,\ell \in \mathbb{Z}_{\geq 0} \text{ and } 0\leq\ell\leq k\ ,
\end{align}
where
\begin{equation*}
\theta_{m,a}(\tau,z) := \sum_{k\in \mathbb{Z}} q^{m(k +\frac{a}{2m})^2} r^{m(k +\frac{a}{2m})}\ .
\end{equation*}
For weights $\tilde{\Lambda} = a \delta + b \alpha_2 + c \delta_N'$ satisfying the condition $\langle \tilde{\Lambda},\delta\rangle <0$,  the character of $\mathcal{B}_N(\slhat)$ when $a=0$ is given by (see \cite[Sec. 3.2]{Govindarajan:2021pkk} for a similar derivation) 
\begin{equation} \label{tildecharacter}
\widetilde{\chi }_{k,\ell }=q^{\frac{1}{8}-\frac{(\ell +1)^2}{4 (k+2)}}\frac{ \chi _{k,\ell }}{T_N(\tau)}\ ,
\end{equation}
with $k=4Nc$ and $\ell=-2b$. The weights are such that $a\in \frac1N\BZ_{\geq0}$ and
$c\in \frac1N\BZ_{>0}$. The character with $a\neq0$ is then $q^a\ \widetilde{\chi }_{k,\ell}$.

\subsection{VVMFs from $\widehat{sl(2)}$ decomposition}

The Jacobi forms $\Psi_{0,m}^{(N)}$ can be expanded in terms of characters of $\slhat$ and those of the Borcherds extension $\mathcal{B}_N(\slhat)$ 
The decomposition takes the form 
\begin{align}
\Psi_{0,m}^{(N)}(\tau,z) &= \sum_{j=-m}^{m} g^{N,m}_{j+1}(\tau)\, \chi_{4m,2m+2j}(\tau,z)\ , \\
&= \sum_{j=-m}^{m} f^{N,m}_{j+1}(\tau)\, \widetilde{\chi}_{4m,2m+2j}(\tau,z)\ ,
\end{align}
Further, one observes that $g^{N,m}_{j+1}(\tau)=g^{N,m}_{-j+1}(\tau)$. This follows from the $\BZ_2$ outer automorphism under which $\alpha_1\leftrightarrow\alpha_2$ and $\alpha_0\leftrightarrow \alpha_3$. Thus one has $(m+1)$ independent functions that we organize into a vector $\mathbf{g}:=(g_1,g_2,\ldots,g_{m+1})^T$. These are rank $(m+1)$ vector valued modular forms of $\Gamma^0{(N)}$.

\noindent \textbf{Remark:} The multiplicities of roots are given the coefficients of the $f_j^{N,m}(\tau)$ which can be obtained from the $g_j^{N,m}(\tau)$ using Eq. \eqref{tildecharacter}.

\subsection{The vvmfs}

Below we give the $\slhat$ decompositions for the $N=2,3,5$ cases. The format is as follows: (i) The vvmf $\mathbf{g}^{N,m}$ has rank $(m+1)$; 
(ii) The first entry of $\mathbf{g}^{N,m}$ is associated with the $\slhat$ character $\chi_{4m,2m}$; (iii) subsequent entries involve a pair of characters related by the $\BZ_2$ automorphism and appear as $\chi_{4m,2m-2j}$ and $\chi_{4m,2m+2j}$ for $j=1,\ldots,m$; (iv) the power of $q$ shown is the one associated with $k=2m,\ell=(2m-2j)$ i.e., $q^{\frac{1}{8}-\frac{(\ell +1)^2}{4 (k+2)}}$ as can be read off from Eq. \eqref{tildecharacter}. We carefully track all roots with positive and zero norm that appear in the character expansion.

\subsubsection{$N=2$}

The coefficients of the Fourier series $T_2(\tau)$ give the mutiplicity of the imaginary simple roots $\delta'_2$ and
 $\delta_2''$. The coefficient of $q^{y}$ gives the multiplicity of the 
 roots $y\,\delta'_2$ and
 $y\,\delta_2''$. One has
\[
T_2(\tau)= 1 \mathbf{- 4}\ q^{1/2} +\mathbf{1}\ q + O(q^{3/2}))\ .
\]
We will see that the expansions below are consistent with these numbers.
\begin{equation*}
\mathbf{g}^{2,1}(\tau)=\begin{pmatrix}
q^{-1/4}(8 q^{1/2}+40 q+128 q^{3/2}+368 q^2+936 q^{5/2}+2176 q^{3}+\cdots)  \\
q^{1/12} (\mathbf{-4}  - 24 q^{1/2} - 88 q - 264 q^{3/2} - 692 q^2 - 1656 q^{5/2} +\cdots)
\end{pmatrix}
\end{equation*}
The leading term in the first row corresponds to the imaginary simple roots $(\alpha^{(1)}_3+\tfrac12 \delta)$ and  $(\alpha^{(1)}_0+\tfrac12 \delta)$ as the constant piece is vanishing. This is consistent with simple real roots $\alpha^{(1)}_3$ and $\alpha^{(1)}_0$ not being present. In the second row, the leading term has multiplicity $-4$ and corresponds to the imaginary roots $\frac12\delta_2'$ and $\frac12\delta_2''$. All other terms correspond to imaginary simple roots with negative norm. 

\begin{equation*}
\mathbf{g}^{2,2}(\tau)=\begin{pmatrix}
q^{-1/2}(-4 q^{1/2} + 2 q - 16 q^{3/2} - 2 q^2 - 56 q^{5/2} + 2 q^3 - 
 144 q^{7/2} +\cdots) \\
 q^{-1/10}(\mathbf{-1} + 4 q^{1/2} + q + 8 q^{3/2} - 2 q^2 + 24 q^{5/2} + 2 q^3 + 
 64 q^{7/2} +\cdots)\\
 q^{1/10} (\mathbf{1} + 8 q^{1/2} + 28 q^{3/2} + 80 q^{5/2} - q^3 + \cdots)
 \end{pmatrix}
 \end{equation*}
 The leading term in the second row above is the multiplicity of the real simple roots $\alpha^{(2)}_0$ and $\alpha^{(2)}_3$. They have multiplicity $1$ and the minus sign comes from $\det(w)$ in the denominator formulae. The Lie algebra $\mathfrak{g}(A^{(2)})$ has four real simple roots. Thus, there are no more simple real roots to track. In the third/last row, the leading term has multiplicity $+1$ and corresponds to the imaginary roots $\delta_2'$ and $\delta_2''$.

\begin{mydef} \label{imagsimpset}
Let $\mathcal{I}$ denote the set of imaginary simple roots with negative norm whose multiplicities  are given by the Fourier expansions of $f_j^{N,m}(\tau)$ for $j=1,\ldots, (m+1)$ and $m=1,\ldots,N$.
\end{mydef}
These are not the complete set of imaginary simple roots as more appear when $m>N$. 

 \subsubsection{$N=3$}
 
The coefficients of the Fourier series $T_2(\tau)$ give the multiplicity of the imaginary simple roots proportional to  $\delta'_3$ and
 $\delta_3''$. The coefficient of $q^{y}$ gives the multiplicity of the 
 roots $y\,\delta'_3$ and
 $y\,\delta_3''$.  One has
\[
T_3(\tau)= 1\mathbf{-3}\ q^{1/3}+ \mathbf{0}\ q^{2/3}\mathbf{-5}\ q+O(q^{4/3}\ .
\]
 
 \begin{equation}
\mathbf{g}^{3,1}(\tau)= \frac{3\eta(\tau)^3}{\eta(\tau/3)^3}\ \begin{pmatrix}
1 \\ -1
\end{pmatrix}= \begin{pmatrix}
q^{-1/4}\ (3\ q^{1/3} + O(q^{2/3}) \\
q^{1/12}\ (\mathbf{-3} + O(q^{1/3})
\end{pmatrix}
\end{equation}
The leading term in the first row corresponds to the imaginary simple roots $(\alpha^{(1)}_3+\tfrac12 \delta)$ and  $(\alpha^{(1)}_0+\tfrac12 \delta)$ as the constant piece is vanishing. This is consistent with simple real roots $\alpha^{(1)}_3$ and $\alpha^{(1)}_0$ not being present. In the second row, the leading term has multiplicity $-4$ and corresponds to the imaginary roots $\frac12\delta_2'$ and $\frac12\delta_2''$. All other terms correspond to imaginary simple roots with negative norm. 

 \begin{equation}
\mathbf{g}^{3,3}(\tau)=  \begin{pmatrix}
q^{-3/4}(14 q + 42 q^{4/3} + 126 q^{5/3} + 308 q^2 + 714 q^{7/3} + 
 1512 q^{8/3} + \cdots) \\
 q^{-9/28}(-3 q^{1/3} - 9 q^{2/3} - 38 q^{1} - 99 q^{4/3} - 252 q^{5/3} - 549 q^{2} + \cdots) \\
q^{-1/28} (-\textbf{1} - 3 q^{1/3} - 9 q^{2/3} - 35 q - 75 q^{4/3} - 180 q^{5/3} - 
 372 q^2 +\cdots) \\
 q^{3/28}(\textbf{5}  + 24 q^{1/3} + 72 q^{2/3} + 191 q + 453 q^{4/3} + 999 q^{5/3} +\cdots)
\end{pmatrix}\notag
\end{equation}

For $N=2,3$, for the terms that we  have studied we are able to see  that the denominator term can be written as
\begin{multline}\label{denominatorterms}
\Delta_{k(N)}^{(N)}(\mathbf{Z}) = \sum_{w\in W} \det(w) w\bigg[(e^{-\varrho}\Big(T_N(\delta) + 
(T_N(\delta'_N)-1) + (T_N(\delta''_N)-1)  \\
+ \sum_{a\in \mathcal{I}} m(a)\ e^{-a} + \cdots \Big)\bigg]
\end{multline}
where the set $\mathcal{I}$ is as defined in Definition \ref{imagsimpset}. The ellipsis refers to contributions from higher orders. Additional terms may be added by incorporating the action of the symmetry $\gamma^{(N)}$ to make the right hand side manifestly invariant under the extended Weyl group. The symmetry under the action of $\widehat{\delta}$ is already present. Terms such as these fit into the Borcherds extension of $\mathfrak{g}(A^{(N)})$. 
 \subsubsection{$N=5$}
 
 The coefficients of the Fourier series $T_5(\tau)$ give the multiplicity of the imaginary simple roots $\delta'_5$ and
 $\delta_5''$. The coefficient of $q^{y}$ gives the multiplicity of the 
 roots $y\,\delta'_5$ and
 $y\,\delta_5''$. 
 One has
\[
T_5(\tau)= 1 \mathbf{-2}\ q^{1/5}\mathbf{-1}\ q^{2/5}+\mathbf{2}\  q^{3/5}+\mathbf{1}\ q^{4/5}+\mathbf{3}\ q+O(q^{6/5})\ .
\]
These appear as the leading coefficient in the bottom row of each vvmf $\mathbf{g}^{5,m}$ for $m=1,\ldots,5$.
 
\begin{equation}
\mathbf{g}^{5,1}(\tau)=\begin{pmatrix}
q^{-1/4}({\color{red} q^{1/5}} + 3 q^{2/5} + 4 q^{3/5} + 7 q^{4/5} + 17 q + 24 q^{6/5} + 
 44 q^{7/5}  +\cdots)  \\
q^{1/12} (-\textbf{2}  - 3 q^{1/5} - 9 q^{2/5} - 12 q^{3/5} - 21 q^{4/5} - 35 q  +\cdots)
\end{pmatrix}\notag
\end{equation} 
For many purposes, it is useful to consider the leading terms in each row. In particular it is easy to extract the weight vector by inspection. In the first row, it is $q^{1/5}\tilde{\chi}_{20,2}$ whose weight vector is $(\frac{\delta}{5}-\alpha_2 + \delta_5')=\frac{\delta}{5}+\alpha^{(5)}_3$ which has norm $2(1-4/5)=2/5$. This is a real fermionic root. Let us call this root $\beta$. Note that $\beta=2\varrho^{(5)}$. One can show that
\begin{equation}
\langle \rho,\beta^{\vee}\rangle = +1\ .
\end{equation}
where the co-root $\beta^\vee :=  \frac{2\beta}{\langle \beta,\beta\rangle}$. Note that this has the `wrong' sign (in our convention) where simple real roots such as $\alpha_1$ are such that $\langle \rho,\alpha_1^{\vee}\rangle = -1$.
 \begin{equation}
\mathbf{g}^{5,2}(\tau)=\begin{pmatrix}
q^{-1/2}({\color{red}q^{2/5}} + q^{3/5} + 2 q^{4/5} + q - 2 q^{6/5} + 7 q^{7/5} + 4 q^{8/5} + 
 8 q^{9/5}+\cdots) \\
 q^{-1/10}(q^{1/5} - 2 q^{2/5} - q^{3/5} - 3 q^{4/5} + q + 5 q^{6/5} - 8 q^{7/5} -  3 q^{8/5} +\cdots)\\
 q^{1/10} (-\textbf{1} - 3 q^{1/5} + q^{2/5} - 2 q^{3/5} - q^{4/5} - 5 q - 12 q^{6/5} + \cdots)
 \end{pmatrix}\notag
 \end{equation}
 \begin{equation}
 \mathbf{g}^{5,3}(\tau)=  \begin{pmatrix}
q^{-3/4}({\color{red}q^{3/5}} + 4 q^{4/5} + 9 q + 14 q^{6/5} + 33 q^{7/5} + 52 q^{8/5} + 
 126 q^{9/5} + \cdots) \\
 q^{-9/28}(-q^{2/5} - 3 q^{3/5} - 15 q^{4/5} - 25 q - 37 q^{6/5} - 74 q^{7/5} - 
 106 q^{8/5}+ \cdots) \\
q^{-1/28} (-3 q^{1/5} - 4 q^{2/5} - 11 q^{3/5} - 2 q^{4/5} - 18 q - 38 q^{6/5} - 
 59 q^{7/5}+\cdots) \\
 q^{3/28}(\textbf{2}  + 9 q^{1/5} + 17 q^{2/5} + 41 q^{3/5} + 53 q^{4/5} + 110 q + 
 201 q^{6/5} +\cdots) \notag
\end{pmatrix}
\end{equation}
\begin{equation}
 \mathbf{g}^{5,4}(\tau)=  \begin{pmatrix}
q^{-1}({\color{red} q^{4/5}} + 2 q + 5 q^{6/5} + 8 q^{7/5} - 2 q^{8/5} + 16 q^{9/5} + 
 13 q^2 + 68 q^{11/5} +\cdots) \\
 q^{-5/9}(2 q^{3/5} - q^{4/5} - 11 q - 11 q^{6/5} + 24 q^{7/5} - 11 q^{8/5} + 
 11 q^{9/5} +\cdots) \\
 q^{-2/9} (-q^{2/5} - 10 q^{3/5} - 7 q^{4/5} - 18 q - 18 q^{7/5} - 103 q^{8/5} - 
 59 q^{9/5} +\cdots)\\
 (-2 q^{1/5} - q^{2/5} + 14 q^{3/5} + 5 q^{4/5} + 19 q - 14 q^{6/5} + 
 6 q^{7/5} + 123 q^{8/5}+\cdots)\\
 q^{1/9}(\textbf{1} + 6 q^{1/5} + 8 q^{2/5} - 6 q^{3/5} + 18 q^{4/5} + 12 q + 
 74 q^{6/5} + 77 q^{7/5} +\cdots)
\end{pmatrix} \notag
\end{equation}
\begin{equation}
 \mathbf{g}^{5,5}(\tau)=  \begin{pmatrix}
q^{-5/4}({\color{red} q} + 2 q^{6/5} + 6 q^{7/5} + 8 q^{8/5} + 14 q^{9/5} - 16 q^2 + 
 40 q^{11/5} + 64 q^{12/5}+\cdots)\\
 q^{-35/44}(5 q - 4 q^{6/5} - 12 q^{7/5} - 16 q^{8/5} - 28 q^{9/5} + 73 q^2 - 
 74 q^{11/5}+\cdots)\\
 q^{-19/44}(-21 q + 6 q^{6/5} + 18 q^{7/5} + 24 q^{8/5} + 42 q^{9/5} - 194 q^2 + 
 112 q^{11/5}+\cdots) \\
 q^{-7/44}(-2 q^{1/5} - 6 q^{2/5} - 8 q^{3/5} - 14 q^{4/5} + 24 q - 40 q^{6/5} - 
 64 q^{7/5}+\cdots) \\
 q^{1/44}(-\textbf{1} + 4 q^{1/5} + 12 q^{2/5} + 16 q^{3/5} + 28 q^{4/5} - 34 q + 
 72 q^{6/5} +\cdots) \\
q^{5/44} (\textbf{3}  - 2 q^{1/5} - 6 q^{2/5} - 8 q^{3/5} - 14 q^{4/5} + 73 q - 
 44 q^{6/5} - 76 q^{7/5}+\cdots)
\end{pmatrix} \notag
\end{equation}
The leading root that appears in $g^{5,5}_4$ is the $\gamma^{(5)}$ image of $q^{1/5}$ and thus appears with the same multiplicity as $q^{1/5}$.
The leading term in the first row of all the vvmfs $\mathbf{g}^{5,m}$ for $m=1,\ldots,5$ is associated with the simple real  root $m\beta$, All appear with multiplicity $+1$ indicating the fermionic nature of the root. These terms are consistent with adding the following term in Eq. \eqref{denominatorterms}.
\begin{equation}\label{nonBorcherdsterm}
\frac{1}{1-e^{-\beta}} = 1 + \sum_{m=1}^\infty e^{-m\beta}\ .
\end{equation}
This is the first term that cannot be a Borcherds correction due to the real nature of the root. However, it would be a Borcherds correction if $\beta$ were a fermionic null root. The first five terms in the above expansion appear in our character expansions with the correct multiplicity. On the product side given by Eq. \eqref{Productside}, we can see that the root $\beta$ appears with  multiplicity $-1$ with the roots $m\beta$ for $m=2\ldots$ not appearing to the extent that we have checked. This is also consistent with the claim in Eq. \eqref{nonBorcherdsterm}.

For $N=2,3$, it expected that $\mathcal{B}^{CHL}_N(A^{(N)})$ is a BKM Lie superalgebra and a suitably enlarged set $\mathcal{I}$ should do the job. As far as we know, an explicit proof is not available in the literature. For $N=5$, we expect a new set of real  roots might appear at $m=10$. In particular, is is known that following two real roots of norm $2$ could appear as they are present in the product side. 
\[
\tilde{\alpha}_{1} =\begin{pmatrix} 4 & 9 \\ 9 & 20 \end{pmatrix} \quad ,\quad
\tilde{\alpha}_{2} = \begin{pmatrix} 6 & 11 \\ 11 &20 \end{pmatrix}\quad.
\]
These are associated with the $\slhat$ characters $\chi_{40,18}$ and $\chi_{40,22}$.  They should appear as the leading coefficient in  the $\slhat$ character decomposition of $\Psi_{0,10}^{(5)}(\mathbf{Z})$ given below. The relevant term in the second row is given in bold face and is vanishing.
\begin{align*}
 \mathbf{g}^{5,10}(\tau)=\begin{pmatrix}
q^{-5/2}({\color{red} q^2}+2 q^{11/5}+5 q^{12/5}+12 q^{13/5}+27 q^{14/5}+114 q^3+\cdots) \\
q^{-85/42}(\mathbf{0\,q^2}+8 q^{11/5}+27 q^{12/5}+20 q^{13/5}+17 q^{14/5}-603 q^3 + \cdots) \\
q^{-67/42}(35 q^2-66 q^{11/5}-207 q^{12/5}-228 q^{13/5}-345 q^{14/5}+\cdots) \\
q^{-17/14}(2 q^{7/5}-8 q^{8/5}-26 q^{9/5}-326 q^2+104 q^{11/5}+461  q^{12/5}+\cdots )\\
q^{-37/42}(5 q-16 q^{6/5}-54 q^{7/5}-40 q^{8/5}-34 q^{9/5}+1056 q^2+\cdots) \\
q^{-25/42}(-35 q+66 q^{6/5}+207 q^{7/5}+228 q^{8/5}+345 q^{9/5}+\cdots )\\
q^{-5/14}(-q^{2/5}+4 q^{3/5}+13 q^{4/5}+164 q-80 q^{6/5}-318 q^{7/5}+\cdots) \\
q^{-1/6}(2 q^{1/5}+9 q^{2/5}-4 q^{3/5}-25 q^{4/5}-397 q+102 q^{6/5}+\cdots )\\
q^{-1/42}(8 q^{1/5}-27 q^{2/5}-20 q^{3/5}-17 q^{4/5}+603 q-352 q^{6/5}+\cdots)\\
q^{1/14}(-3+14q^{1/5}+45 q^{2/5}+44 q^{3/5}+59 q^{4/5}-812 q+\cdots)\\
q^{5/42}(5-16 q^{1/5}-54 q^{2/5}-40q^{3/5}-34 q^{4/5}+1056 q+\cdots )
\end{pmatrix}
\end{align*}
The leading term in row 1 has weight $10\beta$ and multiplicity one. This is consistent with the expansion of the term involving $\beta$ conjectured in Eq. \eqref{nonBorcherdsterm}. The multiplicities are given by the $\mathcal{B}_5^{CHL}(\slhat)$ character expansion. The coefficient of $\widetilde{\chi}_{40,20}$ is
\begin{align*}
f_1^{5,10}(\tau)&= T_5(\tau)(q^2+2 q^{11/5}+5 q^{12/5}+12 q^{13/5}+27 q^{14/5}+114 q^3+ O(q^{16/5} ) \ ,\\
&= q^2+2 q^{13/5}+3 q^{14/5}+63 q^3 + O(q^{16/5})\ .
\end{align*}
The other potential real roots associated with $q^{11/5}$ and $q^{12/5}$ do not appear. 
 
\section{Vector-valued modular forms}

In the previous section, we obtained vector-valued modular forms of the congruence group $\Gamma^0(N)$.  We would like to obtain closed formulae for the Fourier coefficients of these modular forms. In \cite{Govindarajan:2021pkk}, this was done by showing that the vvmfs satisfied a modular differential equation. However, those examples involved modular forms of the full modular group, $PSL(2,\BZ)$.  So we construct vector-valued modular forms for the whole group following a two-step procedure\footnote{We learned this method from the work of Borcherds who obtains modular forms for the full modular group in this fashion\cite{Borcherds:1996uda}. This procedure is called lifting by Bajpai in \cite{Bajpai:2019}.} First, we convert the Jacobi forms of $\Gamma^0(N)$ into Jacobi forms of the full modular group. We obtain vector-valued Jacobi forms in this fashion. Next, we  carry out the character decomposition of the these vector-valued Jacobi forms and obtain vector-valued modular forms of the whole modular group. The price we pay is that the rank of the vector-valued modular forms increases by the index of the subgroup in $PSL(2,\BZ)$.

\subsection{Vector-Valued Jacobi Forms}

The Jacobi forms $\Psi_{0,m}^{(N)}(\tau,z)$ belong to  $J_{0,m}(\Gamma^0(N))$. The Jacobi forms, obtained by the action of $S$, $\psi_{0,m}^{(N)[1,g]}(\tau,z)\Big|S\in J_{0,m}(\Gamma_0(N))$. For prime $N=2,3,5$, there are two cusps of width $1$ and $N$ respectively. We restrict our discussion to only these three cases.  We form a rank $(N+1)$ vector-valued Jacobi Form (vvJF) of the full modular group, $PSL(2,\BZ)$. Let $\psi\equiv \Psi_{0,m}^{(N)}(\tau,z)$ and define
\[
\widetilde{\mathcal{V}}(\psi) =\begin{pmatrix}
\psi(\tau,z)|S \\ \psi(\tau,z) \\  \psi(\tau,z)|T \\ \vdots \\ \psi(\tau,z)|T^{N-1}
\end{pmatrix} \ .
\]
The first entry is the contribution from the cusp at infinity and the other $N$ are the contribution from the cusp at zero. Note that $T^N=1$ at the cusp at zero.
The vvJF, $\widetilde{\mathcal{V}}$, is reducible with $T$ having  an off-diagonal action. We first make a change of basis so that $T$ is diagonal.
 Consider the Jacobi forms (with $\omega_N=\exp(2\pi i/N)$)
\begin{equation}
\widetilde{\psi}_i(\tau,z) = \frac{1}{N}\sum_{j=0}^{N-1} \omega_N^{ij}\ \psi(\tau,z)|T^j\ , \quad i=0,1,\ldots, (N-1)\mod N
\end{equation} 
Now $T$ has a diagonal action i.e., $$\widetilde{\psi}_i(\tau,z)|T =  (\omega_N)^i \ \widetilde{\psi}_i\quad\text{ and }\quad\psi(\tau,z)|ST=\psi(\tau,z)|S\ .$$

The rank $(N+1)$ vvJF $\widetilde{\mathcal{V}}$ is reducible and decomposes into a Jacobi form for the full modular group and another one that is a rank $N$ vvJF. The rank one Jacobi Form is given by the combination
\begin{equation}
\mathcal{A}^{(N)}(\tau,z):=\psi(\tau,z)|S + N \widetilde{\psi}_0(\tau,z)\ .
\end{equation}
and the irreducible rank $N$ vvJF is given by
\begin{equation}
\mathcal{V}^{(N)}(\psi) :=\begin{pmatrix}
\psi(\tau,z)|S - \widetilde{\psi}_0(\tau,z)\phantom{\Big|} \\  \widetilde{\psi}_1(\tau,z) \\ \vdots \\ \widetilde{\psi}_{N-1}(\tau,z)
\end{pmatrix} \ .
\end{equation}
 The $T$ matrix of the vvJF is $$T_V=\text{diag}(1,\omega_N,\ldots, (\omega_N)^{N-1})$$ and the $S$-matrix can be obtained from the following formulae.
\begin{align*}
\left(\psi(\tau,z)|S - \widetilde{\psi}_0(\tau,z)\right)\Big|S &= -\frac{1}{N} \left(\psi(\tau,z)|S - \widetilde{\psi}_0(\tau,z)\right) + \frac{N+1}{N} \sum_{j=1}^{N-1} \widetilde{\psi}_j(\tau,z) \\
\psi(\tau+j,z)|S &= \psi(\tau-j',z) \text{ where } j\neq0\text{ and } jj'=1\text{ mod }N\ .
\end{align*}

For fixed $N$, the S-matrix is independent of the index of the Jacobi form, $\Psi_{0,m}(\tau,z)$ We thus give the S-matrices for the three cases of interest.
\begin{equation}
S_V^{N=2}=\frac12\begin{pmatrix}
  -1 & 3 \\
  1 & 1 
\end{pmatrix}\quad,\quad
S_V^{N=3}=\frac13\begin{pmatrix}
  -1 & 4 & 4 \\
  1 & -1 & 2 \\
  1 & 2 & -1 
\end{pmatrix}
\end{equation}
\begin{equation}
S_V^{N=5}=\frac15\begin{pmatrix}
  -1 & 6 & 6 & 6 & 6 \\
  1 & \frac{1}{2} \left(3-\sqrt{5}\right) & -1-\sqrt{5} & -1+\sqrt{5} & \frac{1}{2} \left(3+\sqrt{5}\right) \\
  1 &-1 -\sqrt{5} & \frac{1}{2} \left(3+\sqrt{5}\right) & \frac{1}{2} \left(3-\sqrt{5}\right) & -1+\sqrt{5} \\
  1 & \sqrt{5}-1 & \frac{1}{2} \left(3-\sqrt{5}\right) & \frac{1}{2} \left(3+\sqrt{5}\right) & -1-\sqrt{5} \\
  1 & \frac{1}{2} \left(3+\sqrt{5}\right) & -1+\sqrt{5} & -1-\sqrt{5} & \frac{1}{2} \left(3-\sqrt{5}\right) \\
\end{pmatrix}
\end{equation}

\subsection{Vector-valued modular forms}

The procedure of the previous sub-section can be applied to all the Jacobi forms, $\Psi_{0,m}^{(N)}(\tau,z)$. In the process we obtain one weight zero modular form that we denote by $\mathcal{A}^{(N)}_m$ and a vvmf of weight zero and rank $N$ that we denote by
$\mathcal{V}^{(N)}_m$ in obvious notation. 

One can decompose the rank $m$ Jacobi form $\mathcal{V}^{(N)}_m$ in terms of $\slhat$ characters, $\chi_{4m,2\ell}$ for $\ell=0,\ldots,2m$ to obtain a rank $(m+1)N$ vector-valued modular form for the full modular group, $PSL(2,\BZ)$. Since the rank grows fast, we will first study the $N=2$ case where we get vvmfs of rank 4 and rank 6. We are able to completely characterize the rank 4 example.
The decomposition is as follows: (with $x=(N-1)(m+1)$)
\begin{equation*}
\mathcal{V}^{(N)}_m = \begin{pmatrix}
g_1\ \chi_{4m,2m} + g_2\ (\chi_{4m,2m-2}+\chi_{4m,2m+2})+ \cdots + g_{m+1}\ (\chi_{4m,0}+\chi_{4m,4m}) \\
\vdots  \\
g_{x+1} \chi_{4m,2m}+ g_{x+2}\ (\chi_{4m,2m-2}+\chi_{4m,2m+2}) +  \cdots +  g_{N(m+1)}(\chi_{4m,0}+\chi_{4m,4m})
\end{pmatrix}\ ,
\end{equation*}
which leads to the vvmf $\mathcal{G}=(g_1,g_2,\ldots,g_{N(m+1)})^T$.
The $S$ and $T$ matrices are, however, easy to write out. Let $S_\chi^{(m)}$ and $T_\chi^{(m)}$  denote the matrices obtained from scalar Jacobi forms of index $m$ as was considered in paper I\cite{Govindarajan:2021pkk}. Then, the S-matrix for the vvmf obtained from $\mathcal{V}^{(N)}_m$ is given by
\begin{equation}
T = T_V^{(N)} \otimes T_\chi^{(m)}\quad\text{and}\quad
S = S_V^{(N)} \otimes S_\chi^{(m)}
\end{equation}
In this fashion, we obtain the data needed to determine the modular differential equation of Gannon\cite{Gannon:2013jua}.

\subsubsection{An example}

Consider $\mathcal{V}^{(2)}_1$ which leads to a rank 4 example. We obtain the following  $T$ and $S$ matrices. 
\begin{equation}
T=\text{diag}\left(e^{-\frac{i\pi}{2}}, e^{\frac{i\pi}{6}} ,e^{\frac{i\pi}{2}}, e^{-\frac{i 5 \pi}{6}}\right) \quad,\quad S=\frac1{2\sqrt3}\begin{pmatrix}
 1 & -2 & -3 & 6 \\
 -1 & -1 & 3 & 3 \\
 -1 & 2 & -1 & 2 \\
 1 & 1 & 1 & 1
\end{pmatrix}
\end{equation}
The first few terms in the Fourier expansion of the vvmf are given below.
\[
\begin{pmatrix}
q^{-1/4} \left(1+36 q+375 q^2+2162 q^3+10017 q^4+38550 q^5+132446 q^6+413478 q^7+\cdots \right) \\
q^{-11/12} \left(-3 q-93 q^2-681 q^3-3723 q^4-15879 q^5-58974 q^6-195186 q^7+\cdots \right)\\
q^{-3/4} \left(-8 q-128 q^2-936 q^3-4784 q^4-19968 q^5-72432 q^6-236392 q^7+\cdots \right)\\
q^{-5/12} \left(24 q+264 q^2+1656 q^3+7848 q^4+31104 q^5+108552 q^6+343992 q^7+\cdots \right)
\end{pmatrix}
\]
Equipped with this data, we can determine the matrix differential equation of Gannon\cite{Gannon:2013jua} to which $\mathcal{G}$ is one of the independent solutions. The data that we need for a rank $d$ situation are the following:
\begin{enumerate}
\item an invertible set of exponents $\Lambda$, and
\item a $d\times d$ matrix $\chi$ defined by
\begin{equation}
\Xi(\tau):=\big(\mathcal{G}_1(\tau),\mathcal{G}_2(\tau),\ldots,\mathcal{G}_d(\tau)\big) = q^\Lambda\ (\mathbf{1}_d + \chi\ q + O(q^2))
\end{equation}
\end{enumerate}
For our rank four example, we obtain
\begin{align}
\Lambda  &=\left(-\frac{1}{4},-\frac{11}{12} ,-\frac{3}{4},-\frac{5}{12}\right) 
\text{ and }\\
\chi &=
\begin{pmatrix}
-8400 & 1296 & 36 & -15876 \\
 72 & 24 & -3 & -32 \\
 -102 & 54 & -8 & 432 \\
 1125 & 106 & 24 & 2800 
 \end{pmatrix}\ .
\end{align}
leading to the four solutions (column 3 is our solution)
\begin{equation*}
{\tiny q^\Lambda=
\begin{pmatrix}
 -8400 q-651744 q^2-17978112 q^3 & 1296 q+28512 q^2+311040 q^3 & 1+36 q+375 q^2+2162 q^3 &
   -15876 q-2094498 q^2-84825468 q^3 \\
 72 q+43056 q^2+2127528 q^3 & 24 q+2064 q^2+33336 q^3 & -3 q-93 q^2-681 q^3 & 1-32 q-50161
   q^2-3921788 q^3 \\
    1-102 q-30051 q^2-1240398 q^3 & 54 q+2268 q^2+33372 q^3 & -8 q-128 q^2-936 q^3 & 432 q+228096
   q^2+14648688 q^3 \\
 1125 q+115650 q^2+3602097 q^3 & 1+106 q+3047 q^2+35814 q^3 & 24 q+264 q^2+1656 q^3 & 2800
   q+518224 q^2+24040112 q^3 \\
\end{pmatrix}
}+ O(q^4)
\end{equation*}
\subsubsection{Other examples}

We are unable to determine the modular differential equations in the other cases. The next lowest rank is six and we need to numerically determine twelve unknown constants. Our attempts to numerically determine the modular differential equation failed. Until rank four, it is easy to determine the modular differential equation making use of an observation of Gannon in \cite{Gannon:2013jua} which enables us to generate three linearly independent solutions given a solution. This puts rank five within reach of numerical computation.

\subsection{The Jacobi Forms $\mathcal{A}_m^{(N)}$}

The $\mathcal{A}_m^{(N)}$ are Jacobi forms for the full modular group. One can expand these as follows:
\begin{equation}
\mathcal{A}_m^{(N)}(\tau,z) = \sum_{j=0}^m h_{2j}(\tau)\ A(\tau,z)^{m-j} B(\tau,z)^j\ ,
\end{equation}
where $h_{2j}(\tau)$ ($j=0,1,\ldots,m$) are modular forms of weight $2j$. Since the ring of modular forms of $PSL(2,\BZ)$ is generated by polynomials in  $E_4(\tau)$ and $E_6(\tau)$, we can characterize $\mathcal{A}_m^{(N)}(\tau,z)$ by a few constants. $h_2(\tau)=0$ since there is no weight two modular form for the full modular group.

In this fashion, we can show that
\begin{align*}
\mathcal{A}_1^{(N)}(\tau,z) &= A(\tau,z) = U_{0,1}(\tau,z) \text{ for } N=2,3,5\ ,\\
\mathcal{A}_2^{(2)}(\tau,z) &= -U_{0,2}\tau,z)\ .
\end{align*}
In these two cases we obtain  Umbral Jacobi forms defined in Eq. \eqref{UJFlist}. That is not true in general. For instance,
\begin{align*}
\mathcal{A}_2^{(3)} &=0\ , \\
\mathcal{A}_3^{(3)} &= \frac1{216} A(\tau,z)^3 +  \frac{5}{72}   E_4(\tau)A(\tau,z) B(\tau,z)^2 -\frac{2}{27} E_6(\tau) B(\tau,z)^3\neq U_{0,3}(\tau,z)\ . 
\end{align*}
$\mathcal{A}_3^{(3)}$ is however a linear combination of two solutions of the matrix differential equation satisfied by the umbral Jacobi form\cite{Govindarajan:2021pkk}.
We are not presenting the Jacobi forms that appear for $N=5$.

\section{Concluding Remarks}

In this paper, we have begun a study of the decomposition of the Siegel modular forms $\Delta^{(N)}_{k(N)}(\mathbf{Z})$ as denominator formulae for a Lie algebra  under two sub-algebras of a Lie algebra, $\mathcal{B}_N^{CHL}(A^{(N)})$, that we wish to understand. There is a natural product formula that provides the product side of the denominator formula -- this provides a description of the positive roots with their multiplicities. The character decomposition that we study is a probe on the sum side of the denominator formula. The work is preliminary as we focused on the first $N$ terms that appear. The $N=5$ case provides the first example of something new. It is the simple real root that we called $\beta$ with $e^{-\beta}\sim q^{1/5}rs$. Roots of type $m\beta$ appear consistent with the expansion of 
\[
\frac{1}{1-e^{-\beta}} = 1 + \sum_{m=1}^\infty e^{-m\beta}\ .
\]
The terms for $m=1,2,3,4,5,10$ that appear in our study agree with the above formula. A preliminary study shows that  similar root with $e^{-\beta}\sim q^{1/6}rs$ appears for the $N=6$ CHL orbifold. We have checked that it again fits the above formula -- we have verified that the first six terms do appear with the correct multiplicity. While the evidence for this is compelling, an all-orders proof is lacking. What is the Lie algebraic interpretation of this kind of `correction' term? There is a  conflict between the following two properties of $\beta$.
\begin{enumerate}
\item The root $\beta$ has positive norm which suggests that it is a real root and should generate a rank one $osp(1,2)$ Lie superalgebra.
\item It appears on the sum side like a Borcherds correction term for an isotropic root. It should generate a rank one $sl(1,2)$ Lie superalgebra.
\end{enumerate}
A resolution of this conflict will go a long way in understanding the Lie superalgebra that we seek.

We also need to work out the cases of $N=4,6$.  The eventual goal is the following: (i) Rewriting the sum term in terms of orbits of the extended Weyl group, (ii) Verifying that the orbits are indeed Borcherds extensions for $N\leq 4$, (iii) For the $N=5,6$ examples, we need to have a good description of \textit{all} terms that don't fit into a Borcherds extension. 

The additive lift for the modular forms $\Delta^{(N)}_{k(N)}(\mathbf{Z})$ was studied in \cite{Govindarajan:2019ezd}. This was done by working out the S-transform of the Hecke operator appearing in an additive lift of Cl\'ery and Gritsenko. This was done for a case by case basis. It would be interesting to carry it out for \textit{all} cases and obtain a closed formula for the sum side. This might enable us to prove that the examples for $N\leq4$ are indeed Borcherds extensions of $\mathfrak{g}(A^{(N)})$.

Our approach to arriving at modular differential equations was blighted by the large ranks that appeared when we constructed vvmfs for the full modular group. The ranks grew as $N(m+1)$ -- the factor of $N$ coming in this process. Is there a way to write modular differential equations for the congruence subgroup? The work of Bajpai might be a way to proceed\cite{Bajpai:2019}. Gottesman has studied rank 2 examples of $\Gamma_0(2)$ in his work\cite{Gottesman:2019}.

\bigskip

\noindent \textbf{Acknowledgements:} We thank S. Samanta and S. Viswanath for collaboration and numerous discussions.
\appendix
\section{Automorphic Forms}
\subsection{Modular Forms}

Let $\mathbb{H}=(\tau ~|~\text{Im}(\tau)>0)$ denote the upper half plane.
\begin{mydef} A modular form, of weight $k$ and character $\chi$, is a function $f:\mathbb{H}\rightarrow \mathbb{C}$ such that for $\gamma=\begin{pmatrix} a& b \\ c& d\end{pmatrix}\in PSL(2,\mathbb{Z})$, one has
\begin{equation}
f|_k \gamma(\tau) = \chi(\gamma)\ f(\tau)\ ,
\end{equation}
where
\[
f|_k \gamma(\tau) := (c\tau + d)^{-k}\ f(\gamma\cdot \tau)\ ,
\]
and $\gamma\cdot \tau = \frac{a\tau+b}{c\tau+d}$.
\end{mydef}
 The level $N$ sub-group $\Gamma_0(N)\subseteq PSL(2,\mathbb{Z})$ comprises those $\gamma$ with $c=0\text{ mod } N$. Similarly, the subgroup $\Gamma^0(N)$ is defined by requiring $b=0\text{ mod } N$.

The group $SL(2,\BZ)$ is generated by two generators that are conventionally called the $T$ and $S$. One has
\[
T:\ \tau \rightarrow \tau+1\quad,\quad S:\ \tau \rightarrow -\frac{1}{\tau}\ .
\]

Let $f(\tau)$ be a modular form of $PSL(2,\BZ)$ with weight $k$. Then, $f(N\tau)$ is a modular form of $\Gamma_0(N)$ and $f(\tau/N)$ is a modular form of $\Gamma^0(N)$ with weight $k$\cite{Atkin:1970}.
Let $j$ be such that $(j,N)=1$. Then we have the following two identities that are very useful.
\begin{equation}\label{identities}
\begin{split}
f\left(\tau/N\right)\big|_kS &= N^k\, f\left(N\tau\right)\\
f\left(\tfrac{\tau+j}{N}\right)\big|_kS &= f\left(\tfrac{\tau-j'}{N}\right)
\end{split}
\end{equation}
with $jj'=1\mod N$. The second line follows from the observation that\cite{Peter2000}
\[
\frac{S\cdot \tau+j}{N} = \frac{j\tau-1}{N\tau}= G \cdot\left(\frac{\tau-j'}{N}\right)\ ,
\]
where $G=\begin{pmatrix}
j & (jj'-1)/N \\ N & j'
\end{pmatrix}\in \Gamma_0(N)$.

\subsubsection{Examples}
A very nice and practical introduction to modular forms is  by Zagier\cite{Zagier:2008}. We define the modular forms that appear in our work.
\begin{enumerate}
\item The Dedekind eta function $\eta(\tau)$  defined by (with $q=\exp(2\pi i \tau)$)
\begin{equation}
\eta(\tau)= q^{1/24} \prod_{m=1}^\infty (1-q^m)\ ,
\end{equation}
is a modular form of weight half and character given by a twenty-fourth root of unity.
\item \textbf{The Eisenstein series:} Let
\begin{align}
E_2(\tau) &= 1 -24 \sum_{n=1}^\infty \sigma_1(n)\ q^n \ , \notag \\
E_4(\tau) &= 1 +240 \sum_{n=1}^\infty \sigma_3(n)\ q^n \ ,\\
E_6(\tau) &= 1 -504 \sum_{n=1}^\infty \sigma_5(n)\ q^n \ . \notag
\end{align}
$E_4(\tau)$ and $E_6(\tau)$ are holomorphic modular forms of $PSL(2,\BZ)$ with weights $4$ and $6$ respectively. They generate the ring of holomorphic modular forms of $PSL(2,\BZ)$. Any holomorphic modular form of $PSL(2,\BZ)$ can be expressed a polynomial of these two modular forms. $E_2(\tau)$ is not modular but 
\[
E_2^*(\tau) = E_2(\tau) -\frac{2}{\text{Im}(\tau)} \ ,
\]
is a non-holmorphic modular form of weight 2.
\item The sub-group $\Gamma_0(N)$ (for $N>1$) has a holomorphic modular form of weight  $2$ given by
\begin{equation}
E_2^{(N)}(\tau) := \frac{1}{N-1} \left(N E_2^*(N\tau)-E_2^*(\tau)\right) =
\frac{1}{N-1} \left(N E_2(N\tau)-E_2(\tau)\right)\ ,
\end{equation}
where we observe that the non-holomorphic pieces cancel away in writing the definition in the second form. It is easy to show that
\[
E_2^{(N)}|S(\tau)=-\frac{1}{N} E_2^{(N)}(\tau/N)\ .
\]
\item Let $\rho=1^{a_1}2^{a_2}\cdots N^{a_N}$ be a cycle shape, for a conjugacy class of $M_{24}$, with $\sum_j j a_j=24$. Then, the product
\[
\eta_\rho(\tau):= \prod_{j=1}^N \eta(j\tau)^{a_j}\ ,
\]
is a modular form $\Gamma_0(N)$ with character given by an $N$-th root of unity\cite{ChengDuncan:2012} (also see \cite{Dummit:1985} for a slightly different version).
\end{enumerate}
\subsection{Ring of Generators for $\Gamma_0(N)$}\label{ring}

Let $M(\Gamma_0(N))$ denote the ring of holomorphic modular forms of $\Gamma_0(N)$. We list the generators of this ring for the cases of interest (obtained from \cite{sage}).
\begin{enumerate}
\item $PSL(2,\BZ)$ has two generators: $E_4(\tau)$ and $E_6(\tau)$.
\item $M(\Gamma_0(2))$ has two generators: $E_2^{(2)}(\tau)$ and $E_4(2\tau)$. 
\item $M(\Gamma_0(3))$ has three generators: $E_2^{(3)}(\tau)$, $E_4(3\tau)$ and $E_6(3\tau)$. 
\item $M(\Gamma_0(5))$ has three generators: $E_2^{(5)}(\tau)$, $E_4(5\tau)$ and $\eta_{1^45^4}=\eta(\tau)^4\eta(5\tau)^4$. 
\end{enumerate}

\subsection{Siegel and Jacobi Forms}

The group $Sp(4,\mathbb{Z})$ is the set of $4\times 4$ matrices written 
in terms of four $2\times 2$ matrices $A$, $B$, $C$,  $D$ (with integral entries)
as
$
M=\left(\begin{smallmatrix}
   A   & B   \\
    C  &  D
\end{smallmatrix}\right)
$
satisfying $ A B^T = B A^T $, $ CD^T=D C^T $ and $ AD^T-BC^T=I $. 
This group acts naturally 
on the Siegel upper half space, $\BH_2$, as
\begin{equation*}
\mathbf{Z}=\begin{pmatrix} \tau & z \\ z & \tau' \end{pmatrix}
\longmapsto M\cdot \mathbf{Z}\equiv (A \mathbf{Z} + B) 
(C\mathbf{Z} + D)^{-1} \ .
\end{equation*}
\begin{mydef}
 A Siegel modular form, of weight $k$ with character $v$ with respect to $Sp(4,\mathbb{Z})$, is a holomorphic function $F:\mathbb{H}_2\rightarrow \mathbb{C} $ satisfying
     \begin{equation}
     F|_k M (\mathbf{Z}) = v(M)\  F(\mathbf{Z})\ ,
     \end{equation}
     for all $M\in Sp(4,\mathbb{Z}) $ where the slash operation is defined as 
 \begin{equation}
    F|_k M(\mathbf{Z}) := \det (C\mathbf{Z}+D)^{-k}\ F(M\cdot \mathbf{Z})\ .
\end{equation}
\end{mydef}

\subsection{Jacobi forms}

In the limit $\tau'\rightarrow i\infty$ or $s=\exp(2\pi i\tau')\rightarrow 0$, a Siegel modular form $\Phi_k(\mathbf{Z})$ has the following Fourier-Jacobi expansion:
\[
\Phi_k(\mathbf{Z}) = \sum_{m=0}^\infty s^m \ \phi_{k,m}(\tau,z) \ .
\]
The Jacobi group $\Gamma_J$ is the sub-group of $Sp(4,\BZ)$ that preserves the condition $s=0$. The transformation of the Fourier-Jacobi coefficients, $\phi_{k,m}(\tau,z)$, under the Jacobi group is a natural definition of a Jacobi form. It is generated by two sub-groups, one is the modular group $PSL(2,\BZ)$ embedded suitably in $Sp(4,\BZ)$ and the other is the Heisenberg group defined below.

The embedding of 
$\left(\begin{smallmatrix} a & b \\ c & d\end{smallmatrix}\right)
\in PSL(2,\BZ)$ in $Sp(4,\BZ)$ is given by
\begin{equation}
\label{sl2embed}
\widetilde{\begin{pmatrix} a & b \\ c & d \end{pmatrix}}
\equiv \begin{pmatrix}
   a   &  0 & b & 0   \\
     0 & 1 & 0 & 0 \\
     c &  0 & d & 0 \\
     0 & 0 & 0 & 1  
\end{pmatrix}
\ .
\end{equation}
The above matrix acts on $\BH_2$ as
\begin{equation}
(\tau,z,\sigma) \longrightarrow \left(\frac{a \tau + b}{c\tau+d},\  
\frac{z}{c\tau+d},\  \sigma-\frac{c z^2}{c \tau+d}\right)\ ,
\end{equation}
with $\det(C\mathbf{Z} + D)=(c\tau +d)$. The Heisenberg group, 
$H(\BZ)$, is generated by $Sp(2,\BZ)$ matrices of the form
\begin{equation}
\label{sl2embedapp}
[\lambda, \mu,\kappa]\equiv \begin{pmatrix}
   1   &  0 & 0 & \mu   \\
    \lambda & 1 & \mu & \kappa \\
     0 &  0 & 1 & -\lambda \\
     0 & 0 & 0 & 1  
\end{pmatrix}
\qquad \textrm{with } \lambda, \mu, \kappa \in \BZ
\end{equation}
The above matrix acts on $\BH_2$ as
\begin{equation}
(\tau,z,\sigma) \longrightarrow \left(\tau,\ z+ \lambda \tau  + \mu,\  
\sigma + \lambda^2 \tau + 2 \lambda z + \lambda \mu +\kappa \right)\ ,
\end{equation}
with $\det(C\mathbf{Z} + D)=1$. 
\begin{mydef}
A Jacobi form of weight $k$ and index $m$ is a map $\phi: \mathbb{H}\times \BZ \rightarrow \mathbb{C}$ satisfying
\[
\Phi|_kM(\mathbf{Z}) = \Phi(\mathbf{Z})\ .
\]
where $\Phi(\mathbf{Z}) := s^m \phi_{k,m}(\tau,z)$. 
\end{mydef}
\noindent The power of $s$ cancels the phases that appear for the Heisenberg group in the usual definition.

\subsubsection{Examples}

The genus-one theta functions are defined by
\begin{equation}
\theta\left[\genfrac{}{}{0pt}{}{a}{b}\right] \left(\tau,z\right)
=\sum_{l \in \BZ} 
q^{\frac12 (l+\frac{a}2)^2}\ 
r^{(l+\frac{a}2)}\ e^{i\pi lb}\ ,
\end{equation}
where $a,\ b\in (0,1)\text{ mod }2$. We define $\theta_1 
\left(\tau,z\right)\equiv i\ \theta\left[\genfrac{}{}{0pt}{}{1}{1}\right](\tau,z)$,
$\theta_2 
\left(\tau,z\right)\equiv\theta\left[\genfrac{}{}{0pt}{}{1}{0}\right] 
\left(z_1,z\right)$, $\theta_3 
\left(\tau,z\right)\equiv\theta\left[\genfrac{}{}{0pt}{}{0}{0}\right] 
\left(\tau,z\right)$ and $\theta_4 
\left(\tau,z\right)\equiv\theta\left[\genfrac{}{}{0pt}{}{0}{1}\right] 
\left(\tau,z\right)$.

The following two index 1 Jacobi forms (with weights 0 and $-2$ respectively) are important.
\begin{align}\label{ABdef}
A_{0,1}(\tau ,z) &=  4\left[ \frac{\theta_2(\tau,z)^2}{\theta_2(\tau,0)^2}+\frac{\theta_3(\tau,z)^2}{\theta_3(\tau,0)^2}+\frac{\theta_4(\tau,z)^2}{\theta_4(\tau,0)^2} \right] = (r^{-1}+10+r) + O(q)\ ,\notag \\
B_{-2,1}(\tau ,z) &=\eta(\tau)^{-6} \theta_1(\tau,z)^2= (r^{-1}-2+r) + O(q)\ .
\end{align}
We usually drop writing the weight and index of these two basic Jacobi forms.
All weak Jacobi forms are given by polynomials in these two Jacobi forms with coefficients given by  modular forms of appropriate weight\cite[see Prop. 6.1]{Aoki:2005}. 

Let $f_i =\theta_i(\tau,z)/\theta_i(\tau,0)$ for $i\in\{2,3,4\}$. The Umbral Jacobi forms at lambency $\ell$ are weak Jacobi forms of weight zero and index $(\ell-1)$\cite{Cheng:2012tq}. We list the three that are relevant for us.
\begin{equation}\label{UJFlist}
\begin{split}
&U_{0,1}(\tau,z) = 4( f_2^2 + f_3^2 + f_4^2)=\left(\tfrac{1}{r}+10 +r\right) + \cdots,\\
&U_{0,2}(\tau,z) = 2(f_2^2 f_3^2 + f_3^2 f_4^2 + f_4^2 f_2^2)=\left(\tfrac{1}{r}+4 +r\right)+\cdots ,\\
& U_{0,3}(\tau,z) = 4 f_2^2 f_3^2 f_4^2=\left(\tfrac{1}{r}+2 +r\right)+\cdots.
\end{split}
\end{equation}

\subsection{Twisted-Twining Elliptic Genera of $K3$}

Let $g$ denote a finite symplectic automorphism of $K3$ of order $N$. We denote one half of the elliptic genus of $K3$ twisted by $g^r$ and twined by $g^s$ by $\psi_{0,1}^{[g^s,g^r]}(\tau,z)$
\begin{equation}
\psi_{0,1}^{[g^s,g^r]}(\tau,z)=\frac{N}{2}\, F_{(N)}^{(r,s)}(\tau,z)\ ,
\end{equation}
where $ F_{(N)}^{(r,s)}(\tau,z)$ are defined in \cite{David:2006ji} for prime $N=2,3,5$ as follows:
\begin{equation}
\label{Frs}
\begin{split}
F_{(N)}^{(0,0)}(\tau ,z)&=\frac{2}{N} A(\tau ,z)  \\
F_{(N)}^{(0,s)}(\tau ,z)&=\frac{2}{N(N+1)}\left[ A(\tau ,z)+ N B(\tau ,z) E^{(N)}_2\left( \tau \right) \right]\quad \text{for}\; 1 \leq s \leq (N-1) \\
F_{(N)}^{(r,r l)}(\tau ,z)&=\frac{2}{N(N+1)}\left[ A(\tau ,z)-  B(\tau ,z)  E^{(N)}_2\left( \tfrac{\tau+l}{N} \right) \right] \\ &\hspace*{2in} \text{for}\; 1 \leq r \leq (N-1),\;0 \leq l \leq (N-1), \notag
\end{split}
\end{equation}

\bibliographystyle{utphys}
\bibliography{refs}

\providecommand{\href}[2]{#2}\begingroup\raggedright\begin{thebibliography}{10}

\bibitem{Govindarajan:2021pkk}
S.~Govindarajan, M.~Shabbir, and S.~Viswanath, ``{$\widehat{sl(2)}$
  decomposition of denominator formulae of some BKM Lie superalgebras},''
  \href{http://dx.doi.org/10.1016/j.nuclphysb.2021.115614}{{\em Nucl. Phys. B}
  {\bfseries 973} (2021) 115614},
  \href{http://arxiv.org/abs/2106.01605}{{\ttfamily arXiv:2106.01605
  [hep-th]}}.

\bibitem{Cheng:2012tq}
M.~C.~N. Cheng, J.~F.~R. Duncan, and J.~A. Harvey, ``{Umbral Moonshine},''
  \href{http://dx.doi.org/10.4310/CNTP.2014.v8.n2.a1}{{\em Commun. Num. Theor.
  Phys.} {\bfseries 08} (2014) 101--242},
\href{http://arxiv.org/abs/1204.2779}{{\ttfamily arXiv:1204.2779 [math.RT]}}.

\bibitem{Govindarajan:2018ted}
S.~Govindarajan and S.~Samanta, ``{Two moonshines for $L_2(11)$ but none for
  $M_{12}$},'' \href{http://dx.doi.org/10.1016/j.nuclphysb.2019.01.004}{{\em
  Nucl. Phys. B} {\bfseries 939} (2019) 566--598},
\href{http://arxiv.org/abs/1804.06677}{{\ttfamily arXiv:1804.06677 [hep-th]}}.

\bibitem{Govindarajan:2019ezd}
S.~Govindarajan and S.~Samanta, ``{BKM Lie superalgebras from counting twisted
  CHL dyons \textendash{} II},''
  \href{http://dx.doi.org/10.1016/j.nuclphysb.2019.114770}{{\em Nucl. Phys. B}
  {\bfseries 948} (2019) 114770},
  \href{http://arxiv.org/abs/1905.06083}{{\ttfamily arXiv:1905.06083
  [hep-th]}}.

\bibitem{Dijkgraaf:1996it}
R.~Dijkgraaf, E.~P. Verlinde, and H.~L. Verlinde, ``{Counting dyons in
  $\mathcal{N}=4$ string theory},''
  \href{http://dx.doi.org/10.1016/S0550-3213(96)00640-2}{{\em Nucl. Phys. B}
  {\bfseries 484} (1997) 543--561},
\href{http://arxiv.org/abs/hep-th/9607026}{{\ttfamily arXiv:hep-th/9607026
  [hep-th]}}.

\bibitem{Cheng:2008kt}
M.~C.~N. Cheng and A.~Dabholkar, ``{Borcherds-Kac-Moody Symmetry of
  $\mathcal{N}=4$ Dyons},''
  \href{http://dx.doi.org/10.4310/CNTP.2009.v3.n1.a2}{{\em Commun. Num. Theor.
  Phys.} {\bfseries 3} (2009) 59--110},
\href{http://arxiv.org/abs/0809.4258}{{\ttfamily arXiv:0809.4258 [hep-th]}}.

\bibitem{Govindarajan:2009qt}
S.~Govindarajan and K.~Gopala~Krishna, ``{BKM Lie superalgebras from dyon
  spectra in $\mathbb{Z}_N$ CHL orbifolds for composite N},''
  \href{http://dx.doi.org/10.1007/JHEP05(2010)014}{{\em JHEP} {\bfseries 05}
  (2010) 014},
\href{http://arxiv.org/abs/0907.1410}{{\ttfamily arXiv:0907.1410 [hep-th]}}.

\bibitem{Govindarajan:2011em}
S.~Govindarajan, ``{Unravelling Mathieu Moonshine},''
  \href{http://dx.doi.org/10.1016/j.nuclphysb.2012.07.005}{{\em Nucl. Phys. B}
  {\bfseries 864} (2012) 823--839},
\href{http://arxiv.org/abs/1106.5715}{{\ttfamily arXiv:1106.5715 [hep-th]}}.

\bibitem{Persson:2013xpa}
D.~Persson and R.~Volpato, ``{Second Quantized Mathieu Moonshine},''
  \href{http://dx.doi.org/10.4310/CNTP.2014.v8.n3.a2}{{\em Commun. Num. Theor.
  Phys.} {\bfseries 08} (2014) 403--509},
  \href{http://arxiv.org/abs/1312.0622}{{\ttfamily arXiv:1312.0622 [hep-th]}}.

\bibitem{David:2006ji}
J.~R. David, D.~P. Jatkar, and A.~Sen, ``{Product representation of Dyon
  partition function in CHL models},''
  \href{http://dx.doi.org/10.1088/1126-6708/2006/06/064}{{\em JHEP} {\bfseries
  06} (2006) 064},
\href{http://arxiv.org/abs/hep-th/0602254}{{\ttfamily arXiv:hep-th/0602254
  [hep-th]}}.

\bibitem{Jatkar:2005bh}
D.~P. Jatkar and A.~Sen, ``{Dyon spectrum in CHL models},''
  \href{http://dx.doi.org/10.1088/1126-6708/2006/04/018}{{\em JHEP} {\bfseries
  04} (2006) 018},
\href{http://arxiv.org/abs/hep-th/0510147}{{\ttfamily arXiv:hep-th/0510147
  [hep-th]}}.

\bibitem{Sen:2007vb}
A.~Sen, ``{Walls of Marginal Stability and Dyon Spectrum in $\mathcal{N}=4$
  Supersymmetric String Theories},''
  \href{http://dx.doi.org/10.1088/1126-6708/2007/05/039}{{\em JHEP} {\bfseries
  05} (2007) 039},
\href{http://arxiv.org/abs/hep-th/0702141}{{\ttfamily arXiv:hep-th/0702141
  [hep-th]}}.

\bibitem{Gritsenko:2002}
V.~A. {Gritsenko} and V.~V. {Nikulin}, ``{On classification of Lorentzian
  Kac-Moody algebras},''
  \href{http://dx.doi.org/10.1070/RM2002v057n05ABEH000553}{{\em Russian
  Mathematical Surveys} {\bfseries 57} no.~5, (Oct, 2002) 921--979},
  \href{http://arxiv.org/abs/math/0201162}{{\ttfamily arXiv:math/0201162
  [math.QA]}}.

\bibitem{Gannon:2013jua}
T.~Gannon, ``{The theory of vector-modular forms for the modular group},''
  \href{http://dx.doi.org/10.1007/978-3-662-43831-2_9}{{\em Contrib. Math.
  Comput. Sci.} {\bfseries 8} (2014) 247--286},
  \href{http://arxiv.org/abs/1310.4458}{{\ttfamily arXiv:1310.4458 [math.NT]}}.

\bibitem{Feingold1983}
A.~J. Feingold and I.~B. Frenkel, ``{A hyperbolic Kac-Moody algebra and the
  theory of Siegel modular forms of genus 2},''
  \href{http://dx.doi.org/10.1007/BF01457086}{{\em Math. Ann.} {\bfseries 263}
  no.~1, (Mar, 1983) 87--144}.
  \url{http://link.springer.com/10.1007/BF01457086}.

\bibitem{Cheng:2008fc}
M.~C.~N. Cheng and E.~P. Verlinde, ``{Wall Crossing, Discrete Attractor Flow,
  and Borcherds Algebra},''
  \href{http://dx.doi.org/10.3842/SIGMA.2008.068}{{\em SIGMA} {\bfseries 4}
  (2008) 068},
\href{http://arxiv.org/abs/0806.2337}{{\ttfamily arXiv:0806.2337 [hep-th]}}.

\bibitem{Kac1990}
V.~G. Kac, {\em Infinite-dimensional {L}ie algebras}.
\newblock Cambridge University Press, Cambridge, third~ed., 1990.

\bibitem{Govindarajan:2020owu}
S.~Govindarajan and S.~Samanta, ``{Mathieu Moonshine and Siegel Modular
  Forms},'' \href{http://dx.doi.org/10.1007/JHEP03(2021)050}{{\em JHEP}
  {\bfseries 03} (2021) 050}, \href{http://arxiv.org/abs/2011.07922}{{\ttfamily
  arXiv:2011.07922 [hep-th]}}.

\bibitem{Clery2008}
F.~Cl\'ery and V.~Gritsenko, ``Siegel modular forms of genus 2 with the
  simplest divisor,'' \href{http://dx.doi.org/10.1112/plms/pdq036}{{\em
  Proceedings of the London Mathematical Society} {\bfseries 102} no.~6, (2011)
  1024--1052}.

\bibitem{Borcherds:1996uda}
R.~E. Borcherds, ``{Automorphic forms with singularities on Grassmannians},''
  \href{http://dx.doi.org/10.1007/s002220050232}{{\em Invent. Math.} {\bfseries
  132} (1998) 491--562},
  \href{http://arxiv.org/abs/alg-geom/9609022}{{\ttfamily
  arXiv:alg-geom/9609022}}.

\bibitem{Bajpai:2019}
J.~Bajpai, ``Lifting of modular forms,'' {\em Publications mathematiques de
  Besancon. Algebre et theorie des nombres} no.~1, (2019) 5--20,
  \href{http://arxiv.org/abs/1705.08363}{{\ttfamily arXiv:1705.08363
  [math.NT]}}.

\bibitem{Gottesman:2019}
R.~Gottesman, ``The arithmetic of vector-valued modular forms on
  {$\Gamma_0(2)$},'' \href{http://dx.doi.org/10.1142/s1793042120500141}{{\em
  International Journal of Number Theory} {\bfseries 16} no.~02, (Sep, 2019)
  241--289}. \url{https://doi.org/10.1142%2Fs1793042120500141}.

\bibitem{Atkin:1970}
A.~O.~L. Atkin and J.~Lehner, ``Hecke operators on {$\Gamma_0(m)$},'' {\em
  Math. Ann.} {\bfseries 185} (1970) 134--160.

\bibitem{Peter2000}
P.~{Niemann}, ``{Some Generalized Kac-Moody Algebras With Known Root
  Multiplicities},'' \href{http://arxiv.org/abs/math/0001029}{{\ttfamily
  arXiv:math/0001029 [math.QA]}}.

\bibitem{Zagier:2008}
D.~Zagier, ``{Elliptic modular forms and their applications},''
  \href{http://dx.doi.org/10.1007/978-3-540-74119-0_1}{{\em The 1-2-3 of
  modular forms} (2008) 1--103}.
  \url{http://link.springer.com/chapter/10.1007/978-3-540-74119-0_1}.

\bibitem{ChengDuncan:2012}
M.~C.~N. {Cheng} and J.~F.~R. {Duncan}, ``{On the Discrete Groups of Mathieu
  Moonshine},'' \href{http://arxiv.org/abs/1212.0906}{{\ttfamily
  arXiv:1212.0906 [math.NT]}}.

\bibitem{Dummit:1985}
D.~Dummit, H.~Kisilevsky, and J.~McKay, ``{Multiplicative products of
  $\eta$-functions},'' {\em Contemp. Math.} {\bfseries 045} (1985) 89--98.

\bibitem{sage}
W.~Stein {\em et~al.}, {\em {S}age {M}athematics {S}oftware ({V}ersion 8.1)}.
\newblock The Sage Development Team, 2017.
\newblock {\tt http://www.sagemath.org}.

\bibitem{Aoki:2005}
H.~Aoki and T.~Ibukiyama, ``Simple graded rings of siegel modular forms,
  differential operators and borcherds products,'' {\em International Journal
  of Mathematics} {\bfseries 16} (03, 2005) 249--279.

\end{thebibliography}\endgroup
\end{document}